\documentclass[11pt]{article}
\usepackage{amsmath}
\usepackage{amsfonts}
\usepackage{amssymb}
\usepackage{graphicx}
\usepackage{geometry}
\usepackage{hyperref}
\usepackage{multirow}
\usepackage{xcolor}
\usepackage[english]{babel}
\usepackage[bottom]{footmisc}
\usepackage{graphicx,algorithmic}
\usepackage[linesnumbered,ruled,vlined]{algorithm2e}
\usepackage[utf8]{inputenc}
\usepackage[T1]{fontenc}
\SetCommentSty{mycommfont}

\SetKwInput{KwInput}{Input}                
\SetKwInput{KwOutput}{Output}              
\title{Computational Taxonomy for visualizing and exploring systems with functional response dynamics: soybean's O2-respiration example.}
\author{Eric Goldman\footnotemark[1], Michelon, Thomas B.\footnotemark[2], and Fushing Hsieh\footnotemark[1]\footnotemark[3]}
\date{today}
\begin{document}
\maketitle
\footnotetext[1]{Department of Statistics, University of California at Davis, CA, 95616. USA.}
\footnotetext[2]{1Department of Plant Science, Federal University of Paran$\acute{a}$, Curitiba, PR, Brazil.}
\footnotetext[3]{Correspondences: Fushing H. Department of Statistics, University of California at Davis, email:fhsieh@ucdavis.edu}

\begin{abstract}
A collective of 1684 O2-respiration curves of 183 dimensions are collected along soybeans' germinating process with 72 covariate feature-measurements. We develop a computational thematic platform for first discovering and exposing heterogeneous typical-shapes and then extracting readable, visible and explainable pattern information with biological implications when comparing between them. Via existing knowledge on seeds of much smaller sizes, such collective of O2-respiration curves are expected to generally embrace the two-stage or sigmoidal patterns. However, due to Soybean's relatively large surface area with harder to control and retain experimental conditions, much diverse and heterogeneous shapes are highly expected. Shannon entropy is applied to capture potential categorical nature across the 183 time-points, while scaled mutual information is used to evaluate pattern-consistency. Three motifs are determined to partition the temporal time-points into three regions: W1 on [1, 20], W2 on [21, 60] and W3 on [61, 183]. Each curve is partitioned into three segments accordingly. Hierarchical cluster algorithm is applied onto these motif-region-specific curve-segments, respectively. Five clusters of shapes on W1 are selected, four on W2 and four on W3. We computationally build a tree-geometry-based Taxonomic Hierarchy to organize these 80 curve-shapes into 15 typical shapes. Then we compare among these typical shapes to extract and display their differences through block-sustained heatmaps. Each focal block is expressed va a network of covariate features to reveal dominant interacting effects of various orders. Scientific explanations and understanding are to be rested on the collected heatmaps and coupled networks collectively.

\end{abstract}

\section{Introduction}
Soybean is one of the most important agricultural products in the world. Two countries: the US and Brazil, produce almost 70\% of the world's total soybean supply. Within this agricultural sector of the two countries and beyond, the recent scientific researches on soybean quality are emerging into a scientific field that is expected to further significantly expand the domain of Seed science. The reason is resting on its relatively large size in comparing with tomato seed and many other plants' seeds. This size characteristic makes it somehow more difficult to be accommodated very well in quality evaluating and experimental settings that are originally designed for seed with small sizes. Further, its large surface area renders higher potential of being contaminated by fungi during its storage process.

Given that the soybean quality categorization technologies have played critical roles in determining the price, it seems natural and necessary to develop robust methodologies that can precisely recognize a spectrum of soybean quality categories when facing the potential of varying degrees of contamination. This research aims at developing such a robust methodology with industrial applicability. Specifically, this methodology is based on data-driven computation, which is free of any man-made structures and assumptions.

This paper reanalyze a data set used in our previous work \cite{michelon}, which was collected along nearly two thousand soybeans' germinating processes. Before discussing our computational developments in this paper, we perform a preliminary data cleaning task to screen out obviously contaminated O2-respiration curves in the original data set. The resultant data set consists 1684 O2-respiration curves with 72 covariate features.

This paper develop a computational thematic platform for first discovering and exposing heterogeneous typical-shapes and secondly extracting and recognizing pattern information when comparing between typical-shapes. That is, we first resolve the issue of how to visualize and discover characteristically distinct shapes embraced by these 1684 functional curves of 183 progressive measurements of remaining O2-contents within a small chamber containing a soybean. We then resolve another issue of how to explain differences between computed and confirmed typical-shapes. In summary, the chief goal of this paper is to provide biologist a comprehensive computational package for categorizing soybean quality. This computational package certainly is applicable to all seeds.

This computing package is termed Computational Taxonomy. By the word ``taxonomy'', this package first compute a Taxonomic Hierarchy with its ending nodes being taken as recognized classes or typical-shapes for O2 respiration curves. This hierarchy presents a heterogeneity-vs-homogeneity map. That is to say that each class is a computed data subset with homogeneity. This computational homogeneity is precisely achieved by declaring that no viable splitting on O2 respiration curves could be supported by information from covariate features. This is a finite-sample phenomenal pattern embraced within a data set.

In contrast, heterogeneity is established as a map of splitting the entire data set. Such a map must be coherent with existing knowledge on seeds. That is, this map will not only create classes exhibiting the two-stage or sigmoidal patterns, but also classes exhibiting other functional-shapes. This capacity is necessary due to Soybean's relatively large sizes. Its large surface area could pose challenges in controlling and retaining ideal experimental conditions. That is why heterogeneity in shape should be highly expected and widely accommodated.

The computations under Taxonomic Hierarchy is based on a developed computing paradigm called Scientific Data Analysis (SDA). Details of SDA can be found in \cite{omotayo,CTonIris,CTonpenguin,CTonCCS}. Basically, all computations are conducted upon the contingency table platform with covariate feature-categories on one-axis, say row-axis, and response-categories on the other axis, say column-axis. One row-vector of a contingency table will sustain one conditional distribution of the response variable. This is a piece of associative information. Since the finite sample randomness pertaining to a contingency table is visible. So, the finite sample precision of each associative information piece is computable and confirmable. We adopt Shannon entropy to evaluate such finite sample precision and to select reliable associative information pieces.

Under anyone response(Re)-covariate(Co) setup, one heatmap will display all computed and confirmed associative information pieces. After properly permuting its row-axis of all selected information pieces and its column-axis of all involving soybean-IDs, an emerging block-structure could demonstrate explicitly the underlying Re-Co dynamics. This is the basis for explanation. We explain each involved component dynamics by a block-specific network, which reveal a collection of high betweenness centrality of hubs and edges. Especially, such edges are isolated, but linked to signal Re-Co dynamics indeed being sustained by high order interacting effects. This discovery is impossible for classic statistical analysis.

From this perspective of discovering and describing dynamics, the merits of Computational Taxonomy is well-illustrated through this categorization of typical-shapes among 1684 O2-respiration curves. Hope seed-biologists and scientists in general find this brand-new data-driven methodology useful for advancing their sciences.

\section{Material and methods}
The collection of 1684 O2-respiration curves with 183 measurements (2/hr) is derived from a data set previously used in \cite{michelon}. All information regarding the experimental set-up and selection criteria for 72 covariate features is referred to the reference. The 72 covariate features includes: 10 ACEs,31 AF, 19 RF, 10 CIE, compactness-cicle, length and Area.  The first four groups of covariate features will undergo fusing to reduce the dimensionality of covariate features. The fusing algorithm would be described below. The total 50 features are fused to become 15 covariate features given as:
\begin{itemize}
\item[1:] $RF_{W1}, RF_{W2}, RF_{W3}, RF_{W4}, RF_{W5}$;
\item[2:] $AF_{W1}, AF_{W2}, AF_{W3}, AF_{W4}$;
\item[3:] $ACE_{W1}, ACE_{W2}$;
\item[4:] $CIE_{bc-mean}, CIE_{bc-sd}, CIE_{Lah-mean}, CIE_{Lah-sd}$.
\end{itemize}
That is, our computing is based on 18 covariate features.

\begin{figure}[ht!]
 \centering
\includegraphics[width=1.0\textwidth]{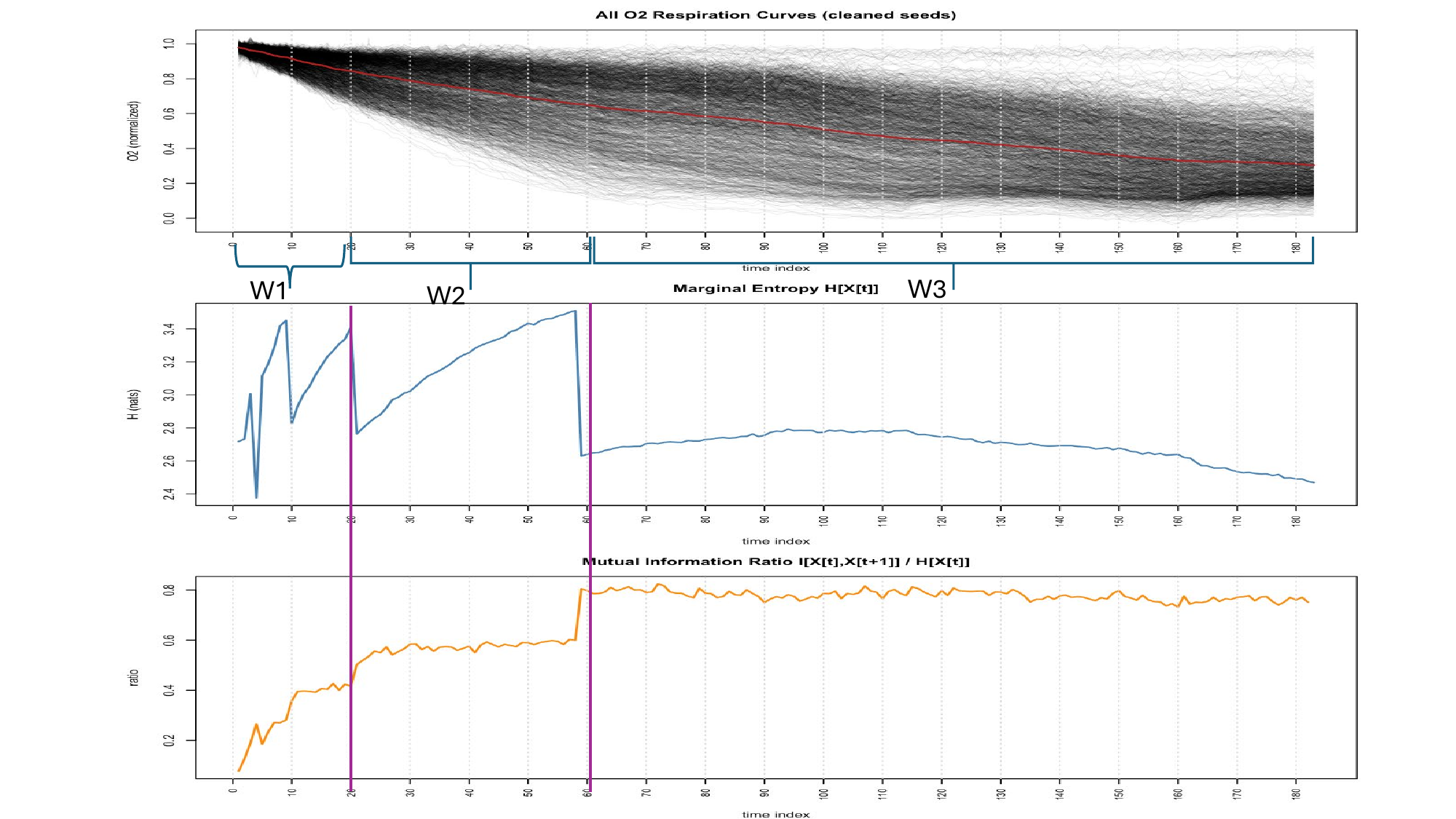}
 \caption{Raw O2 respiration curves (A) coupled with time-point specific entropy curve(B) and mutual information between adjacent time-points (C).}
 \label{O2curves}
 \end{figure}

On the response side, the 1684 functional O2-respiration curves are pooled and plotted with respect to the temporal axis as shown in panel (A) of Fig.~\ref{O2curves}. These O2-respiration curves are tightly tighten on their $(0, 1)$ end. After time origin, these curves' cross-sectional values gradually expanding up to time-point 60, when the maximum variation seems being achieved. Afterward, the amounts of variation continuously decrease only slightly toward time-point 183, when the range of values are widely dispersed across the vertical line.

\begin{figure}[ht!]
 \centering
\includegraphics[width=1.0\textwidth]{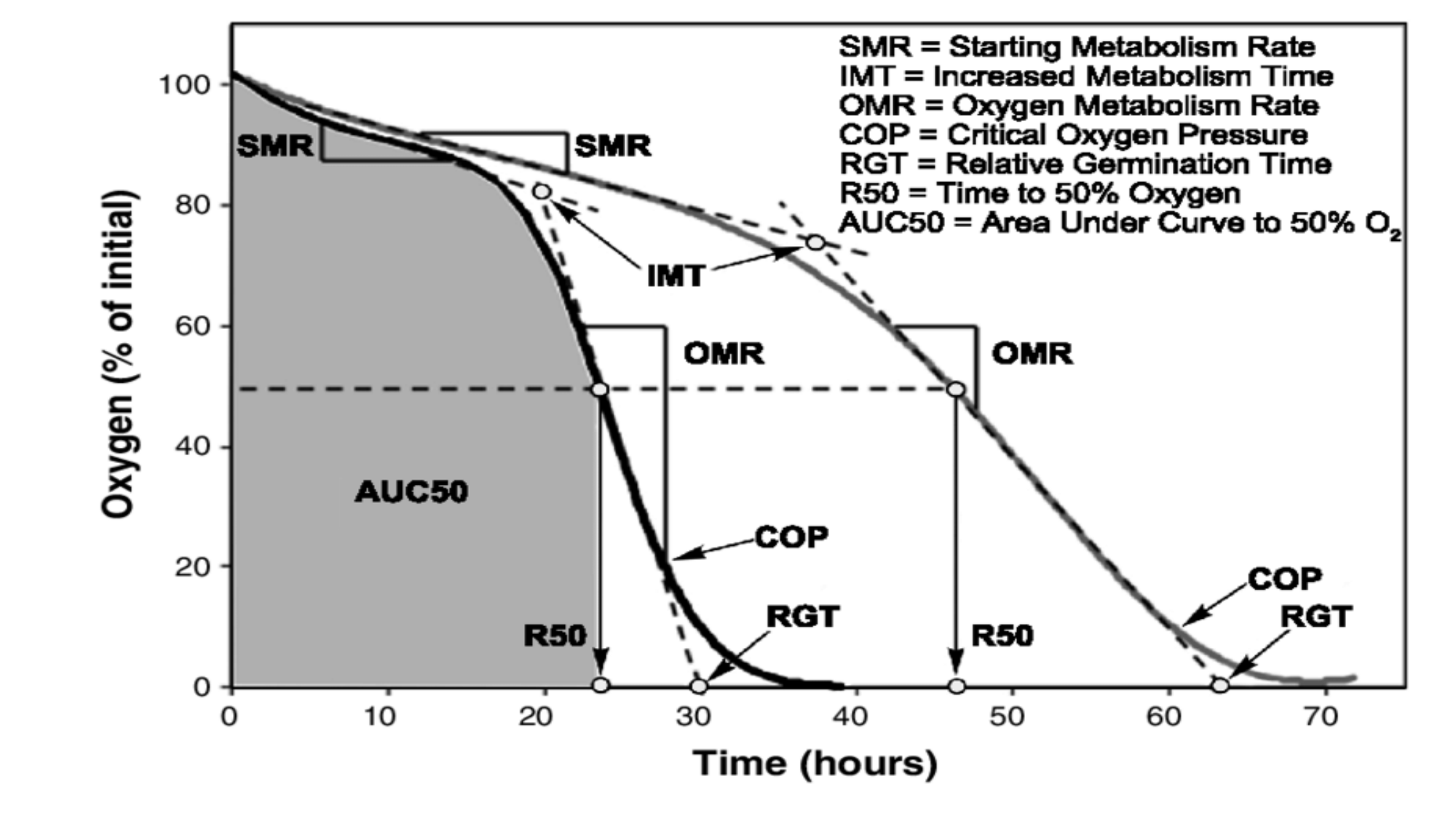}
 \caption{Typical O2 respiration curve with annotated characteristics. Fig.1 of \cite{bradford}.}
 \label{O2Kent}
 \end{figure}

The intensity of this bundle of curves seems to move nonlinearly. This intensity somehow matches the two-stage or sigmoidal patterns described in \cite{bradford}, as shown in Fig.~\ref{O2Kent}.\\
``The typical shape of an oxygen (\(\text{O}_{2}\)) respiration curve during seed germination is a triphasic (three-phased) sigmoidal pattern when plotting cumulative oxygen consumption over time \cite{bradford}.''

It is obvious from panel (A) of Fig.~\ref{O2curves} that only some of 1684 O2-respiration curves follow the sigmoidal curve-patterns. Indeed, many of them don't follow the ideal triphasic (three-phased) sigmoidal pattern. On the other hand, there are several more-or-less visible curve-shapes are recognizable. How many are they? Therefore, discovering the typical-shapes among these 1682 curves is the first task in this data analysis quest.

To tackle this task, it means we need to effectively deal with such visible heterogeneity of curve-shapes. Here, heterogeneity means categorical, not continuity. We select Shannon entropy to accommodate such categorical nature of variations across the entire temporal span. Specifically, we build a Hierarchical Clustering (HC) tree upon all O2-values, and cut this HC-tree at $K (?)$ clusters. Then, we adopt this $k$-clusters as one categorization scheme to categorize all O2-values. As such, a categorical variable $X(t)$ is observed at time-point $t$. We then calculate Shannon entropy $H[X(t)]$ for all 183 time-points, as shown in panel (B) of Fig.~\ref{O2curves}. This 1D entropy curve reveal gradual-up and then sudden-drop patterns at 10, 20 and 60 time-points. Afterward, the entropy curve goes rather steady along the temporal-axis beyond time-point 60.

Next, we calculate re-scaled mutual information $I[X(t), X(t+1)]/H[X(t)]$, where mutual information $I[X(t), X(t+1)]$ is calculated as:
\[
I[X(t), X(t+1)]= H[X(t)]-H[X(t)|X(t+1)]].
\]
The re-scaled mutual information $I[X(t), X(t+1)]/H[X(t)]$ evaluates the proportion of entropy $H[X(t)]$ being ``explained'' by ``knowing'' the information of $X(t+1)$. This evaluation is used to measure the degree of pattern-consistency across the entire temporal span. The curve of $I[X(t), X(t+1)]/H[X(t)]$ is shown in panel (C) of Fig.~\ref{O2curves}. This re-scaled mutual information curve definitely indicates that three temporal regions: $[0, 20]$,$[21, 60]$ and $[61, 183]$, embrace rather characteristically distinct varying patterns of the 1684 O2-respiration curves.

\begin{figure}[ht!]
 \centering
\includegraphics[width=1.0\textwidth]{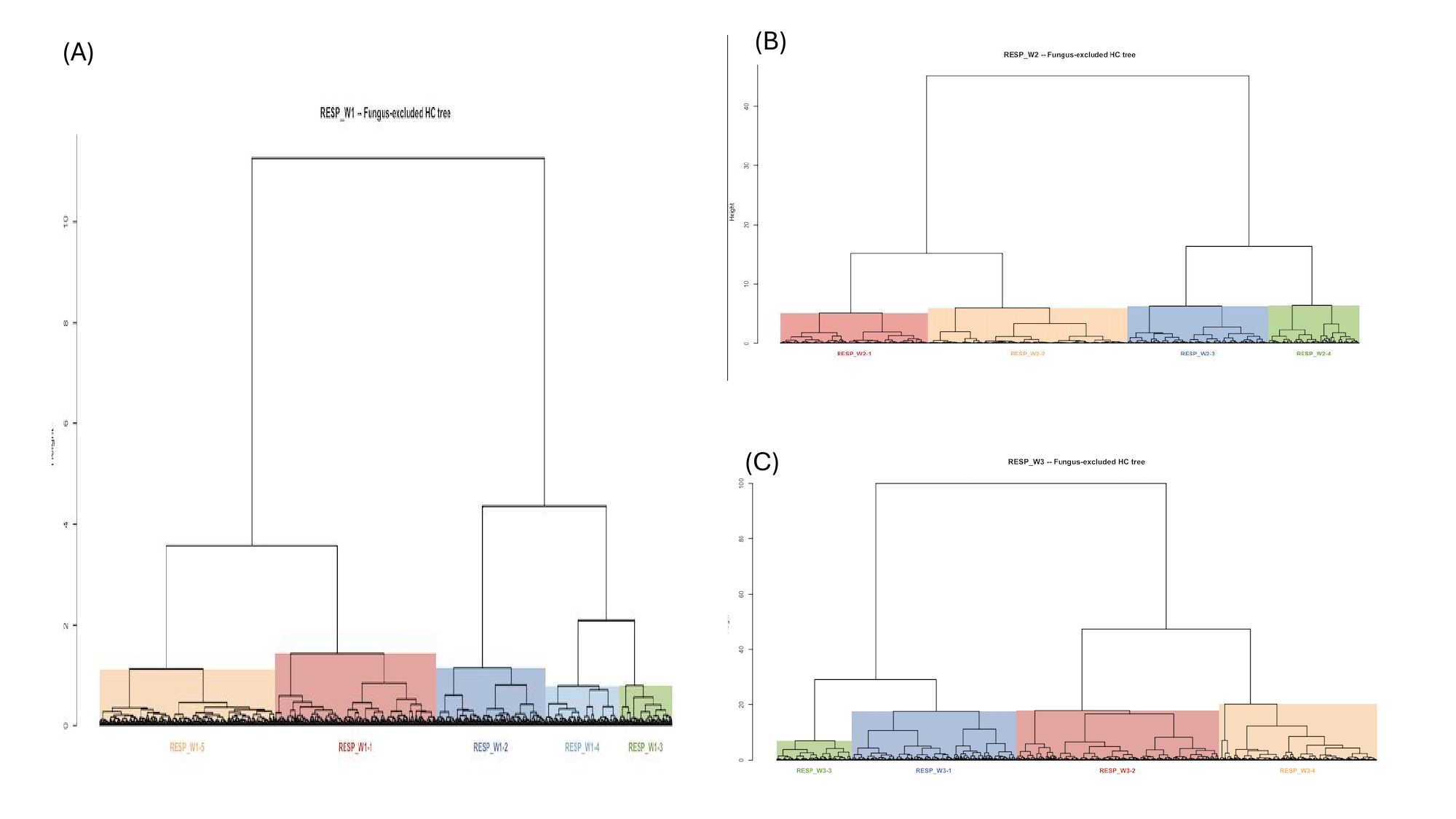}
 \caption{Three HC-tree of $W1, W2, W3$.}
 \label{HCtreeW123}
 \end{figure}

\begin{figure}[ht!]
 \centering
\includegraphics[width=1.0\textwidth]{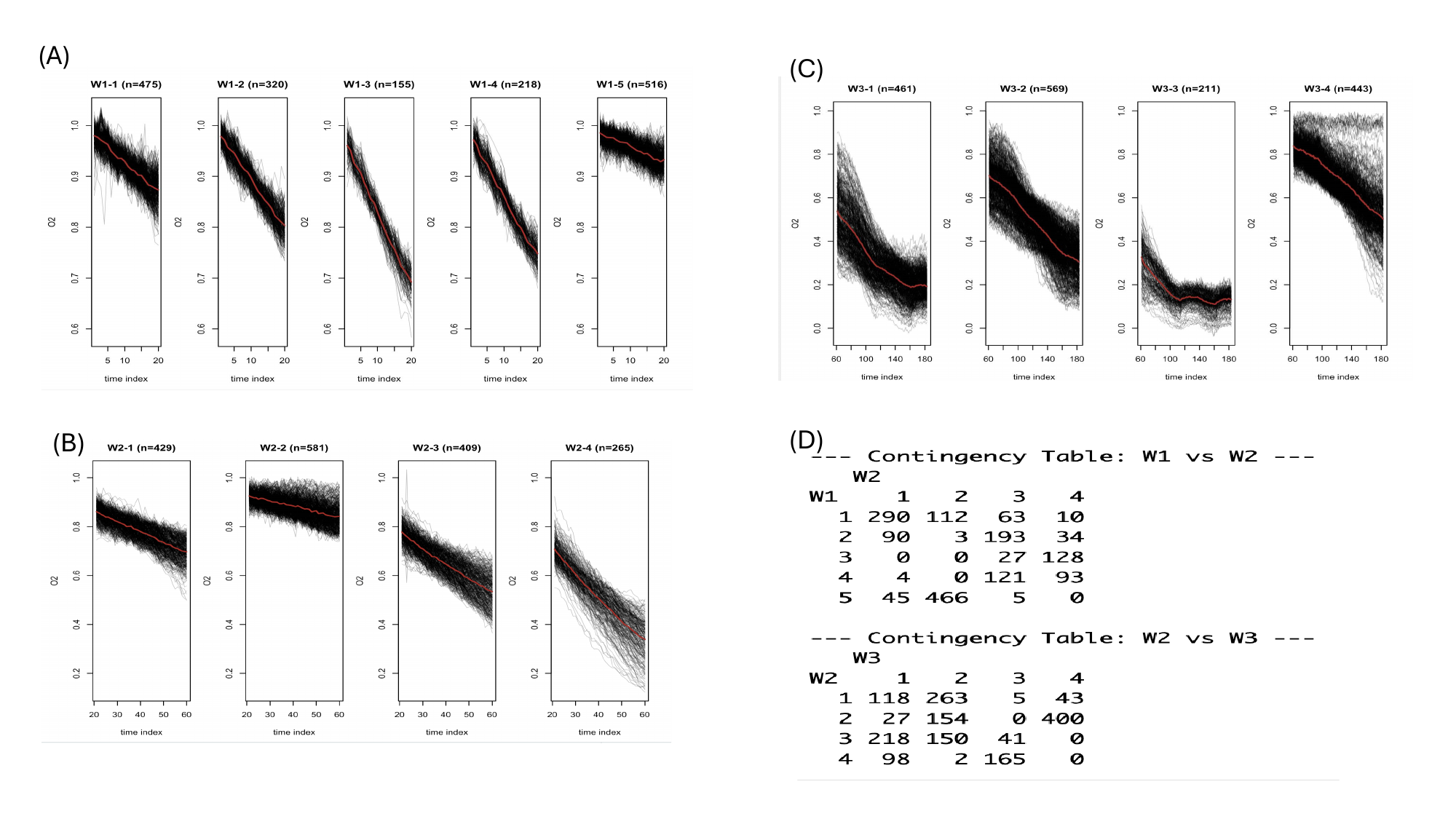}
 \caption{Shapes of functional curve-segments of $W1, W2, W3$.}
 \label{shapeW123}
 \end{figure}

Accordingly, we partition each O2-respiration curve into three segments and denote them as: W1, W2 and W3. That is, W1 is a segment of 20 measurements, W2 is a segment of 39 and W3 is a segment of 123 O2-measurements. Then we apply HC algorithm with Euclidean distance and Ward-D2 module onto these three collectives of 1684 segments, respectively. The three HC-trees with marked clusters are reported in Fig.~\ref{HCtreeW123}. The collection of 1684 W1 segments are clustered into 5 clusters, both W2 and W3 segments into 4 clusters, respectively. Their cluster-specific segments are displayed in Fig.~\ref{shapeW123}.

\begin{figure}[ht!]
 \centering
\includegraphics[width=1.0\textwidth]{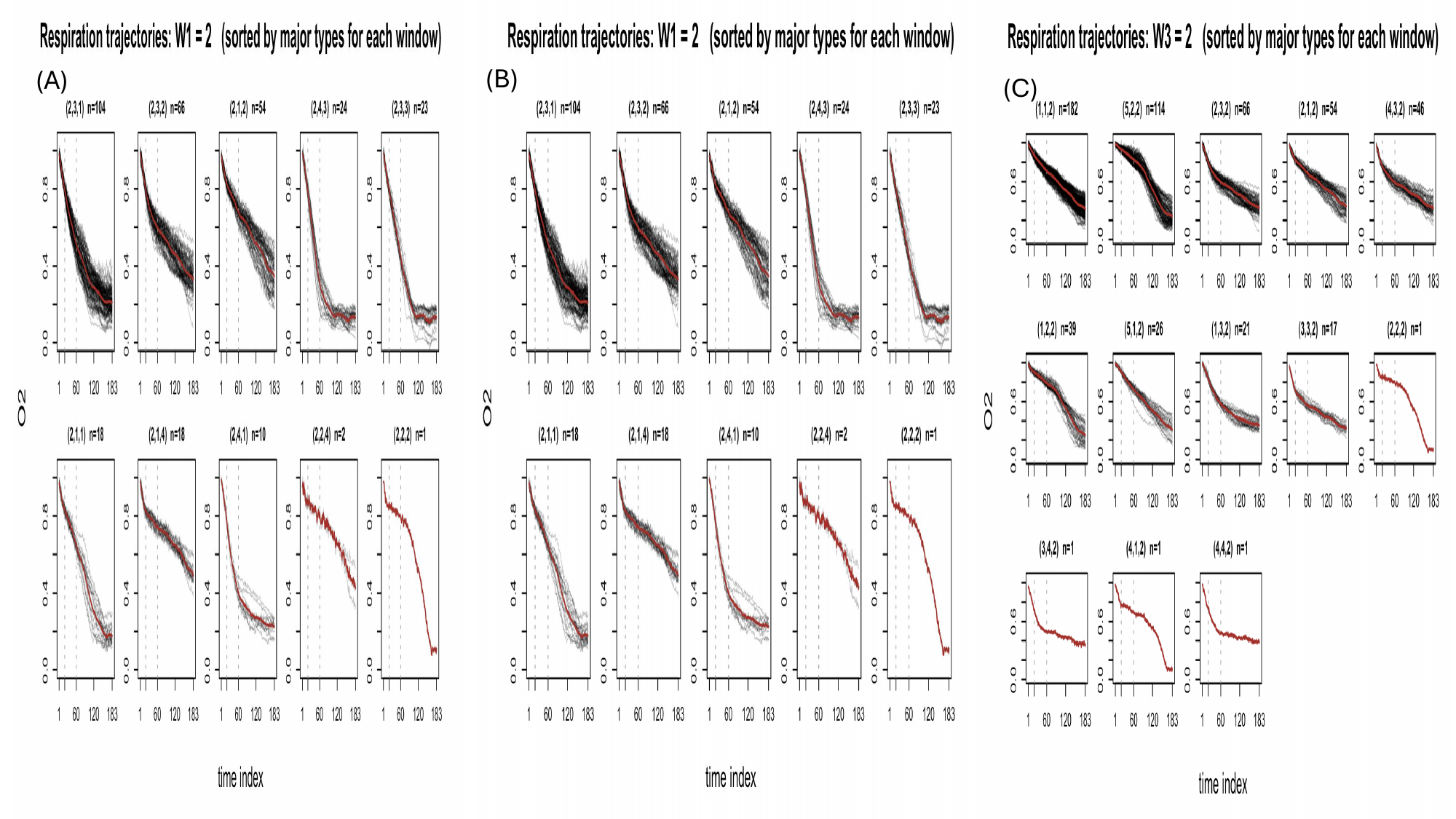}
 \caption{Shapes of whole functional curve-segments of selected $W1, W2, W3$.}
 \label{FullshapeW123}
 \end{figure}

Three glimpses of whole curve-view of a cluster of W1, W2 and W3, respectively, are presented in Fig.~\ref {FullshapeW123}. All of these cells of O2-respiration curves are relatively ``slender'' along their mean-curves. However, many of them take rather similar shapes. Further, there could be as many as $80(=5\times 4\times 4)$ curve-shapes via all possible combinations of clusters of W1, W2 and W3. This number: 80, is certainly too many due to the shape-similarity seen in Fig.~\ref {FullshapeW123}. How to re-organize them effectively without compromising the shape-typicality? We address this issue in the next section.

Before the end of this section, we discuss briefly the fusing scheme on the covariate side. For ordered covariate features: ACEs, AFs and RFs, we perform the Shannon entropy and re-scaled mutual information calculations like what has been done on the response side. We make respective grouping decisions on ordered covariate features: 1) two for ACEs; 2) 4 for AFs; 3) 5 for RFs. Upon each group of covariate features, we apply HC-algorithm to fuse all group-member features into one categorical variables with a number of categories resulted from cutting its HC-tree. For 10 CIEs, we also apply the same protocol of grouping and fusing to arrive at four categorical variables: $CIE_{bc-mean}, CIE_{bc-sd}, CIE_{Lah-mean}, CIE_{Lah-sd}$.

\section{Key shape visualization}
It is like a simple question: How many typical shapes of leaves can be found within a tree? With its simplicity in description, this scientific question is by-no-means simple. Even the intuitive ideas about how to think about it is still missing in literature. Not mentioning actually resolve it? The reason can be attributed to the key character of system: Heterogeneity-vs-homogeneity. Heterogeneity is not a mathematical term, while homogeneity is hardly well-defined in real-world systems. Here homogeneity would be taken as a typical-shape of O2-respiration curves.

How many typical-shapes can be computed within a large collection of O2 respiration curves? By selecting 5 clusters for fused categorical variable W1, 4 for W2 and W3, then we have less than $80=(5\times 4\times 4)$ with respect to triplets of cluster-IDs of $(W1, W2, W3)$. Though some of the non-empty triplet-combinations have relative large numbers, majority of them have relatively small numbers. These small-size triplet combinations of $(W1, W2, W3)$ are unlikely typical. How to reorganize among these combinations to achieve a much smaller set of typical-shapes. In this section, a hierarchical approach is proposed. The resultant hierarchy is called Taxonomic Hierarchy.

The first step of constructing a Taxonomic hierarchy upon all triplets of cluster-IDs of $(W1, W2, W3)$ is to recognize dependence through two contingency tables among these three categorical variables as shown in panel (D) of Fig.~\ref {FullshapeW123}. The contingency table of W1-vs-W2 contains several zero-cells, so does the contingency table of W2-vs-W3. Intuitively and indeed naturally, such dependence is shape-dependence oriented. In other words, the HC-trees of W1, W2 and W3 are dependent.

How to explore and see patterns of dependence among the three HC-trees of W1, W2 and W3? This is the main task taken up in this section. We build Taxonomic Hierarchy to reveal the span of patterns of HC-tree-dependence as one whole in a step-by-step fashion. Most importantly, we demonstrate that such HC-tree-dependence patterns must be justified via covariate information. As such the map of Heterogeneity-vs-homogeneity emerges.

\subsection{1st layer of Taxonomic Hierarchy.}
Upon the whole 1684 O2-respiration curves, what is the basis for making the first split? Given that HC-tree of W1, W2 and W3 are all binary tree. At first, we determine splitting W1, W2 and W3 into binary response variables via similarity-grouping:1) W1(1\&5)-vs-W1(2\&3\&4);2)W2(1\&2)-vs-W2(3\&4);
3)W3(1\&3)-vs-W3(2\&4), according to their HC-tree structures, which are coherent with their curve-shapes across their categories.

\begin{figure}[ht!]
 \centering
\includegraphics[width=1.0\textwidth]{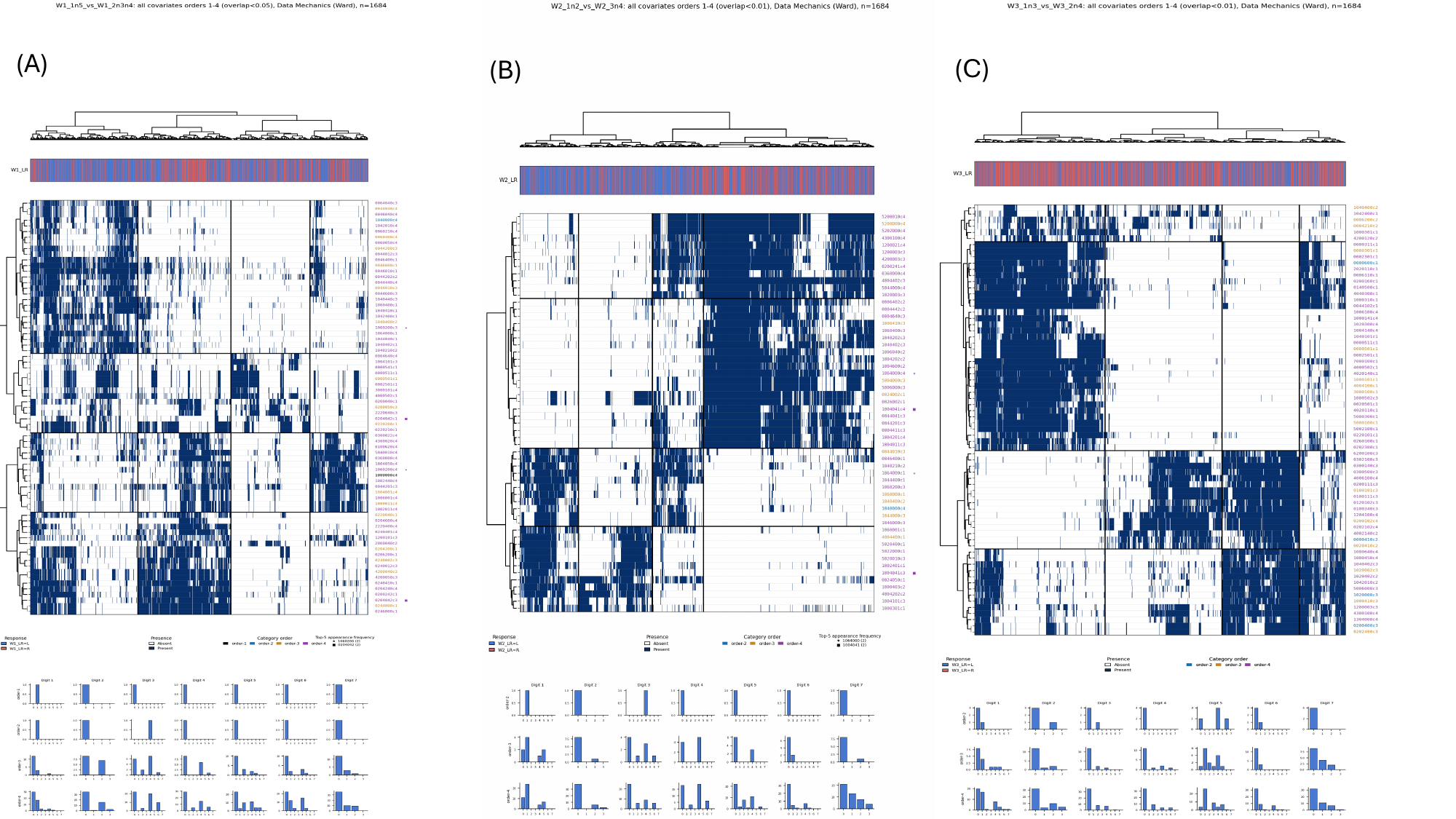}
 \caption{Three heatmaps for (A) W1(1\&5)-vs-W1(2\&3\&4); (B)W2(1\&2)-vs-W2(3\&4); (C)W3(1\&3)-vs-W3(2\&4)}
 \label{Ttree1st}
 \end{figure}

The criterion of determining which binary split to start is based on the choice of which binary split is most supported by covariate information. That is, the split: W1(1\&5)-vs-W1(2\&3\&4), is taken as binary response variable. What are all pieces of associative information provided by the 15 covariate features? We apply the recently developed Scientific Data Analysis (SDA) computing paradigm \cite{omotayo,CTonIris,CTonpenguin,CTonCCS} to computed and confirm ``reliable'' associative information pieces. All such associative information pieces are displayed on a heatmap platform as shown in panel (A) of Fig.~\ref{Ttree1st}. Likewise, a SDA based heatmap for splitting W2(1\&2)-vs-W2(3\&4) is given in panel (B), and a heatmap for splitting W3(1\&3)-vs-W3(2\&4) is panel (C).

Comparing the three heatmap visually, we see that the blue-vs-red contrast is more evident in panel (B) than that of panels (A) and (C). The split: W1(1\&5)-vs-W1(2\&3\&4), is somehow blur, while the splitting of W3(1\&3)-vs-W3(2\&4) is rather unbalanced because of the dominance of the red. As we choose the binary split: W2(1\&2)-vs-W2(3\&4), the first layer of Taxonomic Hierarchy is done with two branches: L1 for W2(1\&2) and R1 for W2(3\&4). It is essential to see that the heatmap in panel (B) shows signs of heterogeneity by having two separating primary blue clusters.

\subsection{2nd layer}
\begin{figure}[ht!]
 \centering
\includegraphics[width=1.0\textwidth]{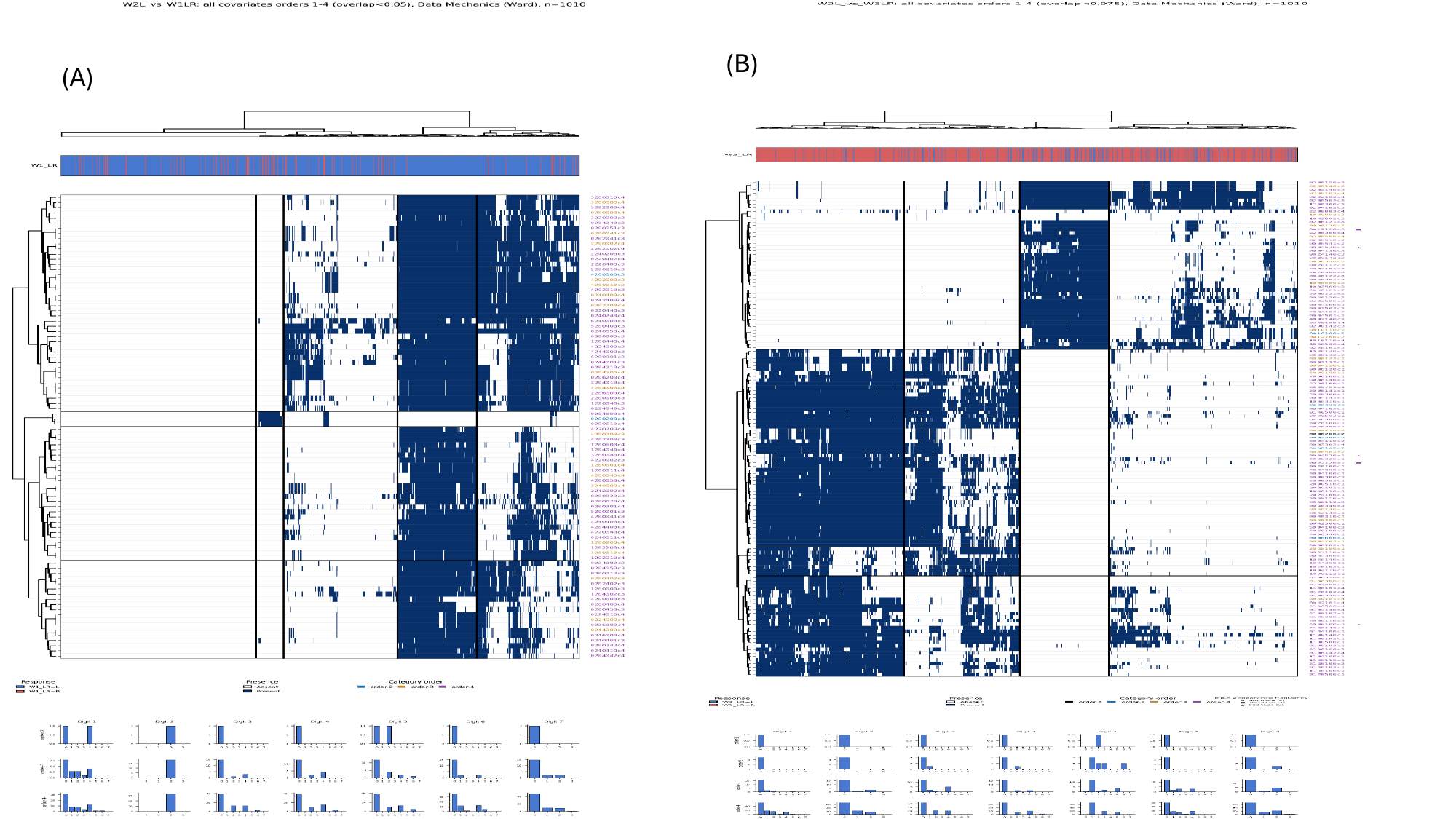}
 \caption{Two heatmaps for splitting L1(=W2(1\&2)) w.r.t (A) W1(1\&5)-vs-W1(2\&3\&4); (B)W3(1\&3)-vs-W3(2\&4)}
 \label{Ttree2ndL1}
 \end{figure}

We then go on to build the 2nd layer by splitting its left branch (L1): W2(1\&2), and right branch (R1):W2(3\&4), with respect to 1) W1(1\&5)-vs-W1(2\&3\&4);2)W3(1\&3)-vs-W3(2\&4), respectively. Again by applying SDA computing, two heatmaps for splitting W2(1\&2) as L1 are shown in two panels of Fig.~\ref{Ttree2ndL1}. Based on comparing the two heatmaps, the split of W1(1\&5)-vs-W1(2\&3\&4) in panel (A) is picked. As such branch L1 is split into L1L2 and L1R2. Here branch L1L2 is for W1(1\&5), while L1R2 is for W1(2\&3\&4). Again the heatmap in panel (A) shows signs of heterogeneity by having separating primary blue clusters. Therefore, further splitting is needed.

\begin{figure}[ht!]
 \centering
\includegraphics[width=1.0\textwidth]{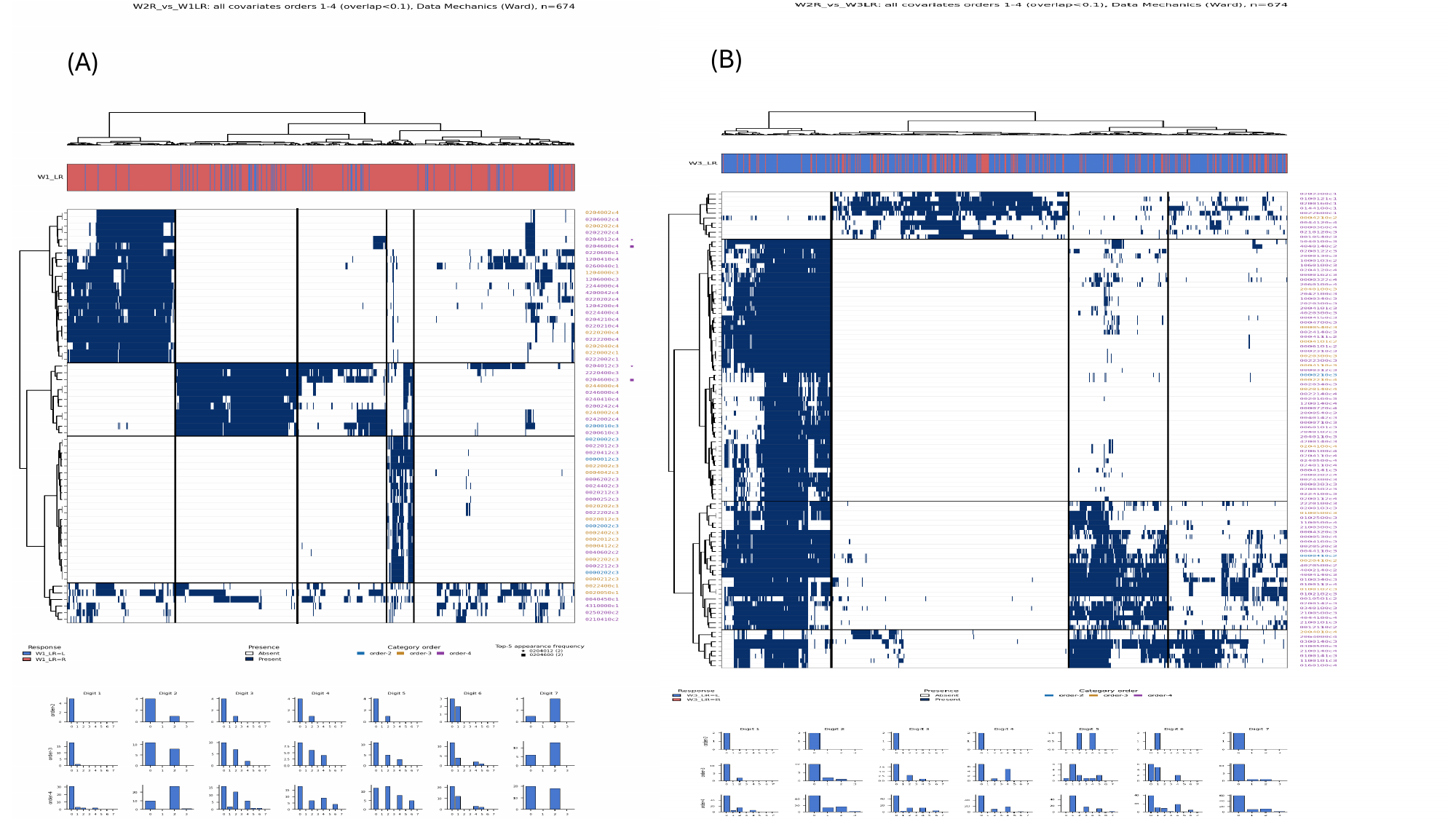}
 \caption{Two heatmaps for splitting L1(=W2(3\&4)) w.r.t (A) W1(1\&5)-vs-W1(2\&3\&4); (B)W3(1\&3)-vs-W3(2\&4)}
 \label{Ttree2ndR1}
 \end{figure}

By applying SDA computing, two heatmaps for splitting W2(3\&4) as R1 are shown in two panels of Fig.~\ref{Ttree2ndR1}. And the split of W1(1\&5)-vs-W1(2\&3\&4) is picked. As such branch R1 is split into R1L2 and R1R2. Here branch R1L2 is for W1(1\&5), while R1R2 is for W1(2\&3\&4).

So, the 2nd layer of Taxonomic Hierarchy is completed with four branches: L1L2, L1R2 on the left as subbranches of L1 and R1L2 and R1R2 on the right as subbranches of R1. Likewise the heatmap in panel (A) also shows signs of heterogeneity by having separating primary red clusters. Therefore, further splitting is also needed.

\subsection{3rd layer}
\begin{figure}[ht!]
 \centering
\includegraphics[width=1.0\textwidth]{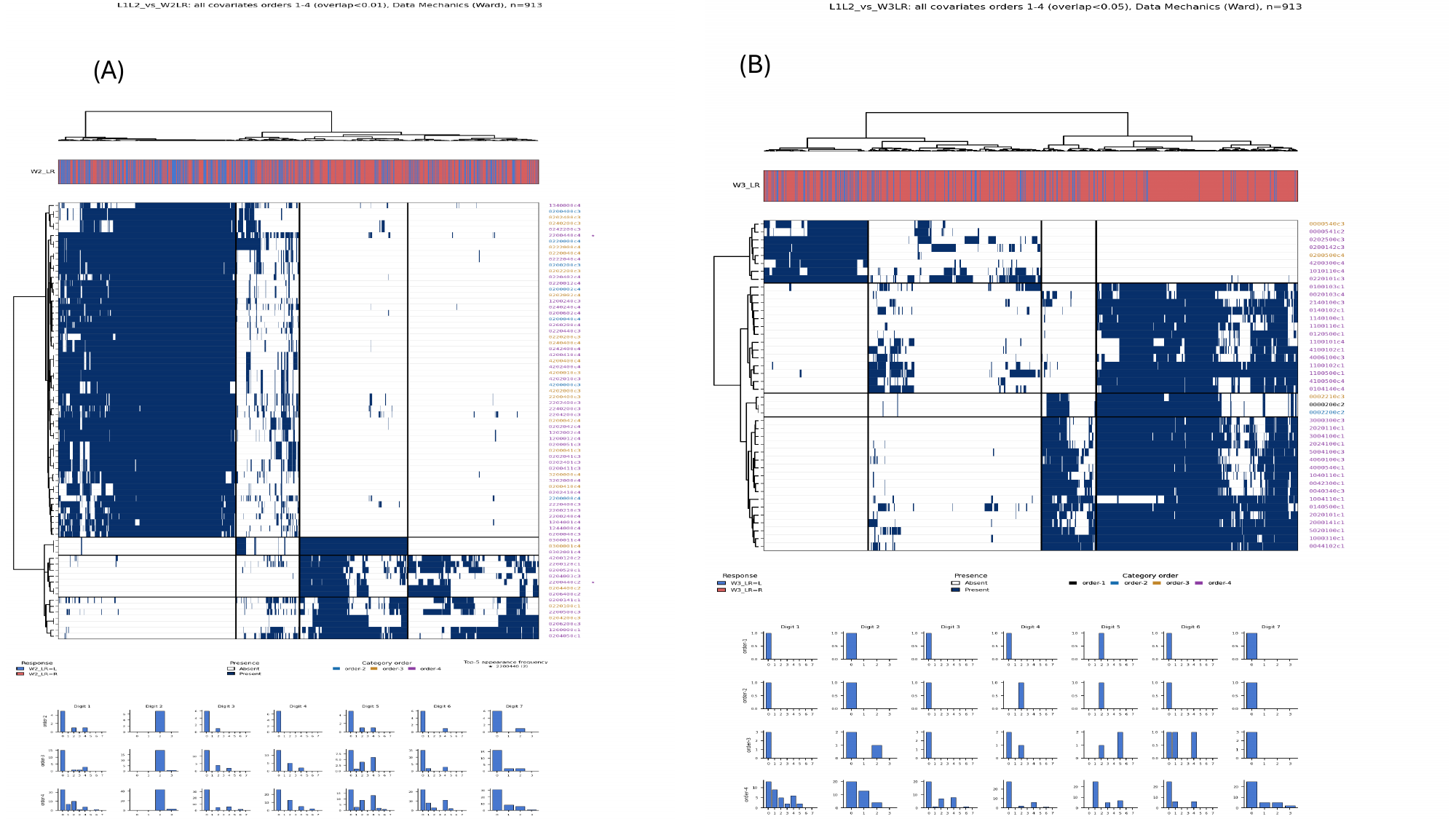}
 \caption{Two heatmaps for splitting L1l2 w.r.t (A) W2(1)-vs-W2(2); (B)W3(1\&3)-vs-W3(2\&4)}
 \label{Ttree3rdL1L2}
 \end{figure}

On the 3rd layer, two heatmaps for splitting L1L2 is performed and compared with respect to:(1) W2(1)-vs-W2(2); (2) W3(1\&3)-vs-W3(2\&4). Both heatmaps in (A) and (B) panels of Fig.~\ref{Ttree3rdL1L2} show signs of heterogeneity. They signs are seemingly equally strong. Since splitting with respect to W3 hasn't been selected in the previous two layers. So, we select the splitting by W3(1\&3)-vs-W3(2\&4) to result into two sub-branches: L1L2L3 and L1L2R3.  We will further split L1L2L3 and L1L2R3 in the 4th layer.

\begin{figure}[ht!]
 \centering
\includegraphics[width=1.0\textwidth]{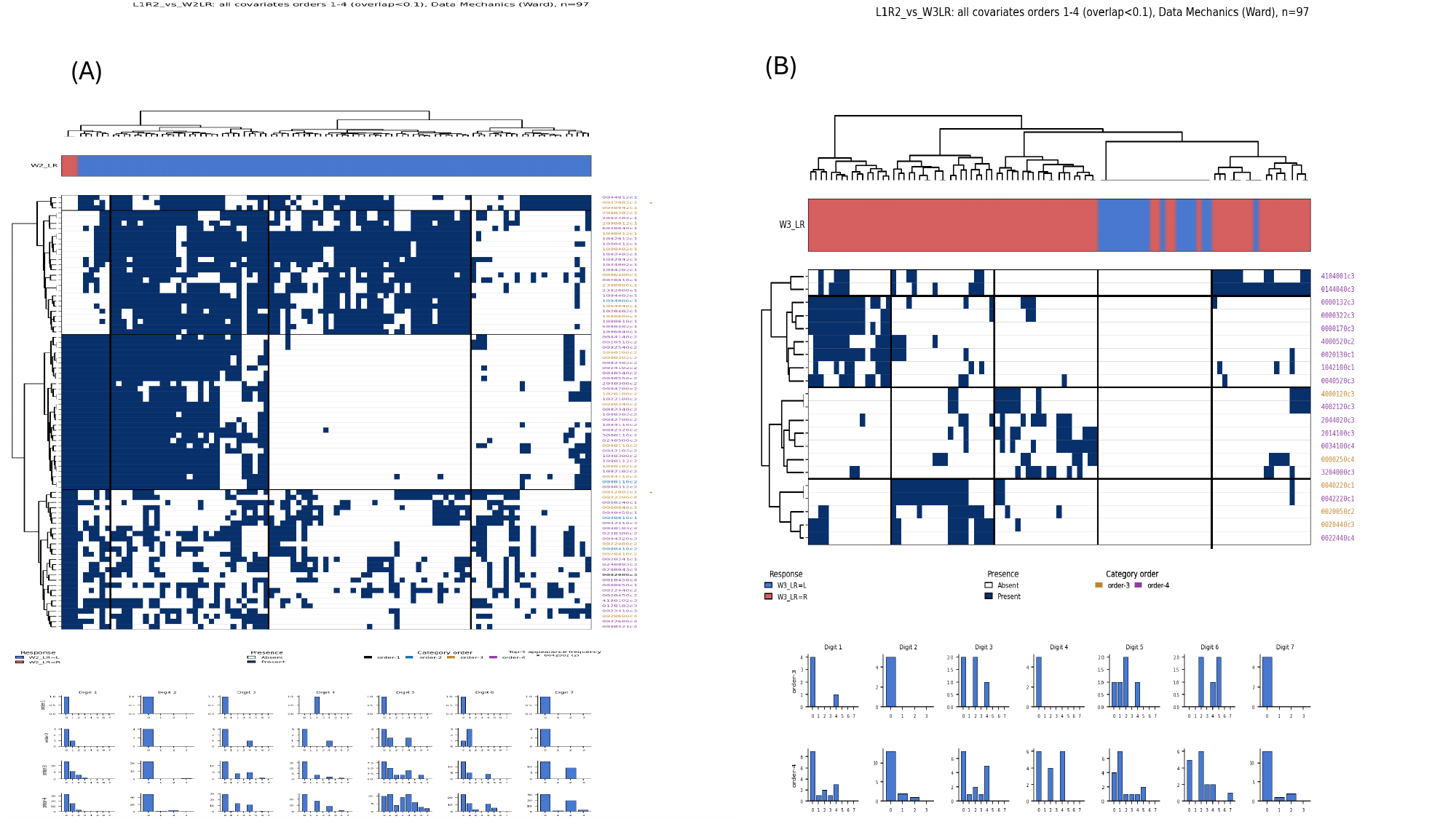}
 \caption{Two heatmaps for splitting L1R2 w.r.t (A) W2(1)-vs-W2(2); (B)W3(1\&3)-vs-W3(2\&4)}
 \label{Ttree3rdL1R2}
 \end{figure}

Next, two heatmaps for splitting L1R2 is performed and compared with respect to:(1) W2(1)-vs-W2(2); (2) W3(1\&3)-vs-W3(2\&4). The heatmap in (B) panel of Fig.~\ref{Ttree3rdL1R2} show strong signs of heterogeneity on red, but not the one in panel (A). Therefore, we select the splitting by W3(1\&3)-vs-W3(2\&4) to result into two sub-branches: L1R2L3 and L1R2R3. Both sizes of both sub-branches become small. Hence, we decide no further splitting on both sub-branches. That is, L1R2L3 and L1R2R3 become two ending nodes within the Taxonomic Hierarchy.

\begin{figure}[ht!]
 \centering
\includegraphics[width=1.0\textwidth]{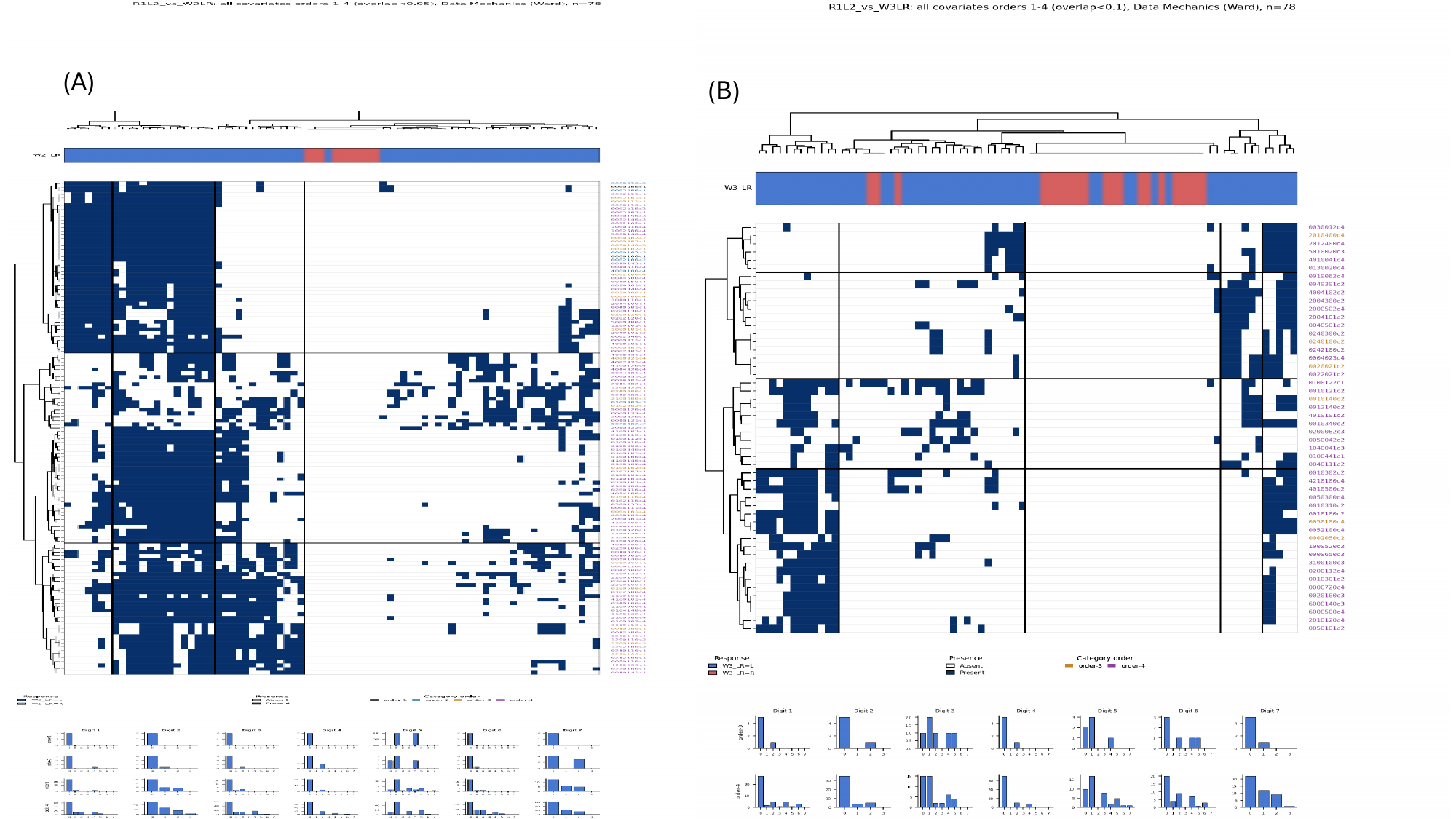}
 \caption{Two heatmaps for splitting R1L2 w.r.t (A) W2(3)-vs-W2(4); (B)W3(1\&3)-vs-W3(2\&4)}
 \label{Ttree3rdR1L2}
 \end{figure}

Thirdly, two heatmaps for splitting R1L2 is performed and compared with respect to:(1) W2(1)-vs-W2(2); (2) W3(1\&3)-vs-W3(2\&4). The heatmap in (B) panel of Fig.~\ref{Ttree3rdR1L2} show strong signs of heterogeneity than the one in panel (A). Therefore, we again select the splitting by W3(1\&3)-vs-W3(2\&4) to result into two sub-branches: R1L2L3 and R1L2R3. Both sizes of both sub-branches also become small. Hence, we decide no further splitting on both sub-branches. That is, R1L2L3 and R1L2R3 also become two ending nodes within the Taxonomic Hierarchy.

\begin{figure}[ht!]
 \centering
\includegraphics[width=1.0\textwidth]{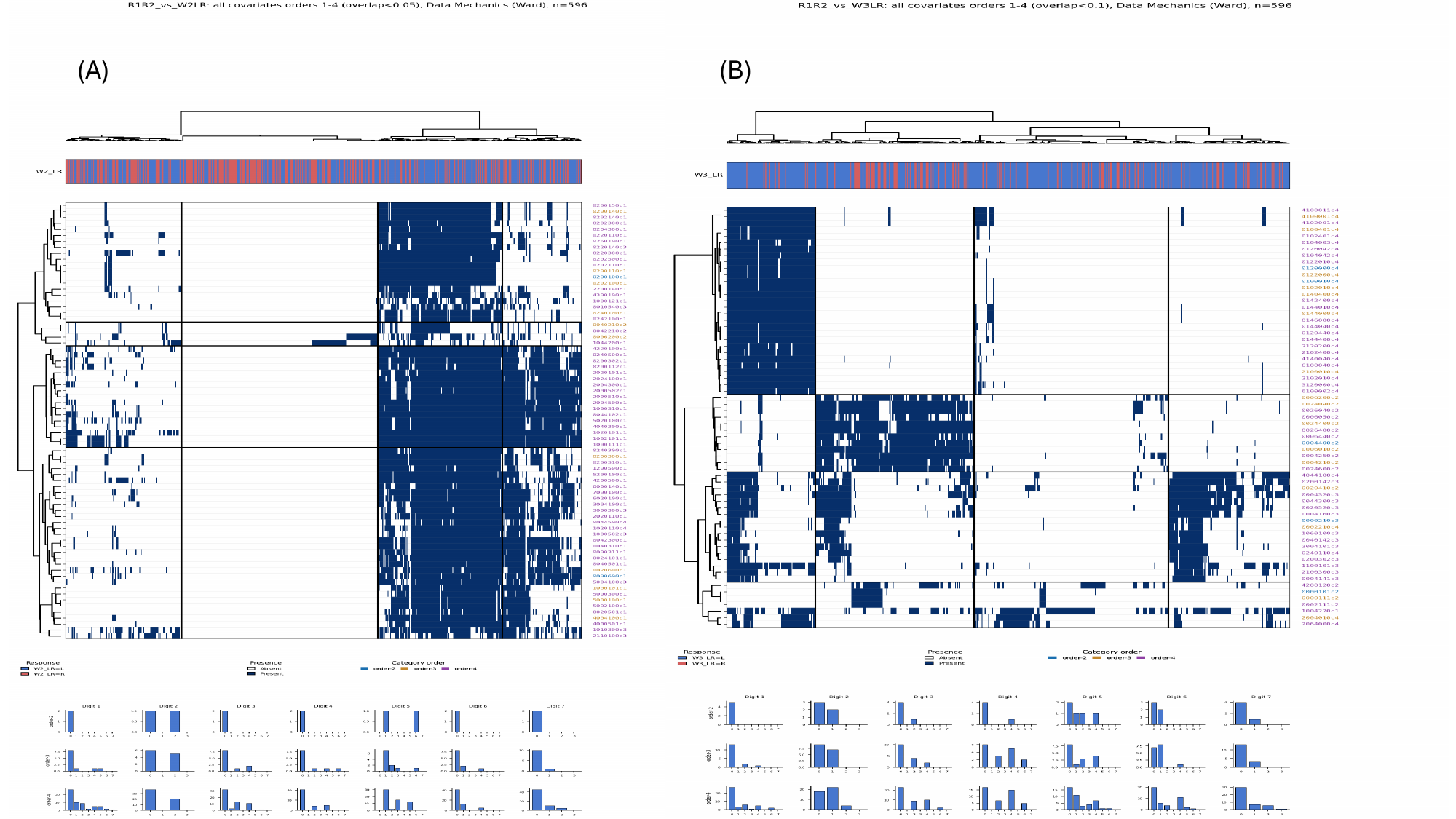}
 \caption{Two heatmaps for splitting R1R2 w.r.t (A) W2(3)-vs-W2(4); (B)W3(1\&3)-vs-W3(2\&4)}
 \label{Ttree3rdR1R2}
 \end{figure}

Fourthly, two heatmaps for splitting R1R2 is performed and compared with respect to:(1) W2(1)-vs-W2(2); (2) W3(1\&3)-vs-W3(2\&4). The heatmap in (B) panel of Fig.~\ref{Ttree3rdR1R2} show strong signs of heterogeneity than the one in panel (A). Therefore, we select the splitting by W3(1\&3)-vs-W3(2\&4) to result into two sub-branches: R1R2L3 and R1R2R3. Both sizes of both sub-branches also become small. Hence, we decide no further splitting on both sub-branches. We will further split R1R2L3 and R1R2R3 in the 4th layer.

\subsection{4th layer}
\begin{figure}[ht!]
 \centering
\includegraphics[width=1.0\textwidth]{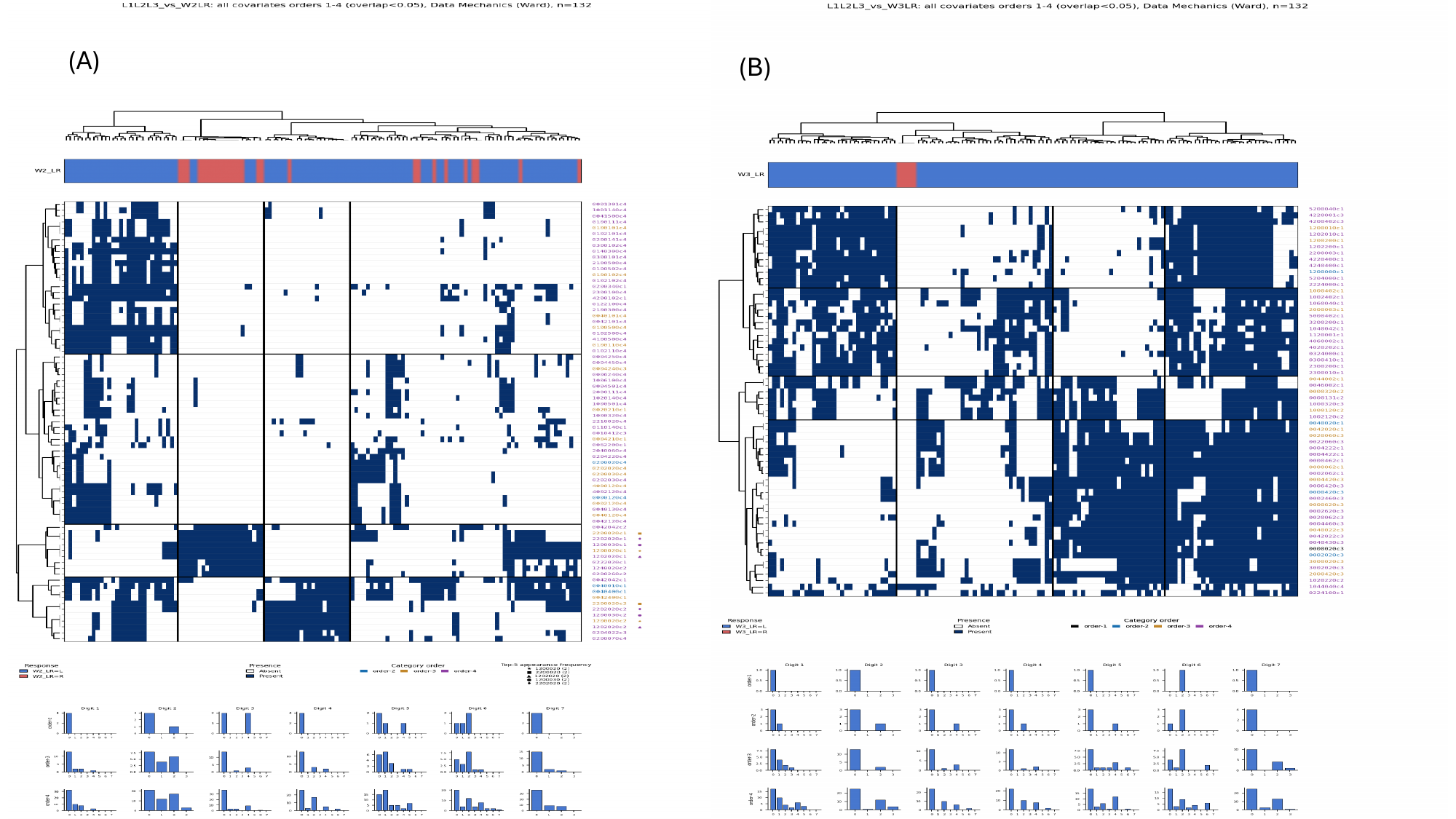}
 \caption{Two heatmaps for splitting L1L2L3 w.r.t (A) W2(1)-vs-W2(2); (B)W3(1)-vs-W3(3)}
 \label{Ttree4thL1L2L3}
 \end{figure}

\begin{figure}[ht!]
 \centering
\includegraphics[width=1.0\textwidth]{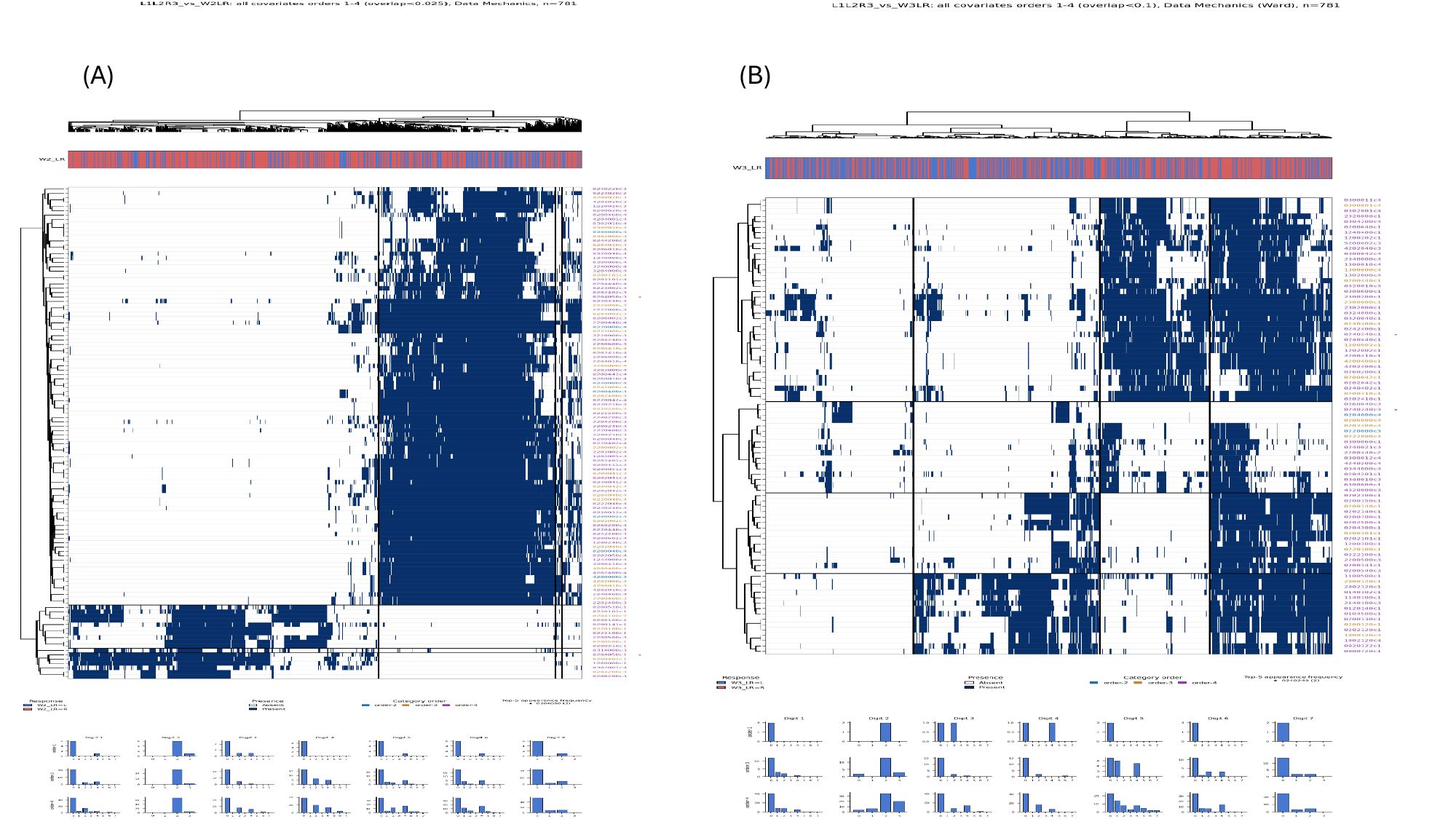}
 \caption{Two heatmaps for splitting L1L2R3 w.r.t (A) W2(1)-vs-W2(2); (B)W3(2)-vs-W3(4)}
 \label{Ttree4thL1L2R3}
 \end{figure}

Upon the 4th layer of Taxonomic Hierarchy, we first split L1L2L3 with respect to: 1)W2(1)-vs-W2(2); 2) W3(1)-vs-W3(3). The heatmap in panel (A) of Fig.~\ref{Ttree4thL1L2L3} show clear patterns separating blues and reds, while panel (B) is a heatmap with very small number of reds. So it is clear to choose the splitting by W2(1)-vs-W2(2) to generate two sub-branches: L1L2L3L4 and L1L2L3R4. Further, the heterogeneity among blues is not clear within this heatmap even though two clusters of blues being present. So, we can guess that L1L2L3L4 and L1L2L3R4 are likely two ending nodes. But this speculation needs supporting evidence.
Next, we split L1L2R3 with respect to: 1)W2(1)-vs-W2(2); 2) W3(2)-vs-W3(4). Both heatmaps in the two panels of Fig.~\ref{Ttree4thL1L2R3} seem to show evident, but complex heterogeneity due to the fact of involving with large number soybeans. That is, further splitting will be required afterward. Therefore, we choose to split via W2(1)-vs-W2(2) first on this layer. Two sub-branches: L1L2R3L4 and L1L2R3R4, are resulted. Likely, both branches are to be split with respect to W3(2)-vs-W3(4) in the 5th layer of Taxonomic Hierarchy.

\begin{figure}[ht!]
 \centering
\includegraphics[width=1.0\textwidth]{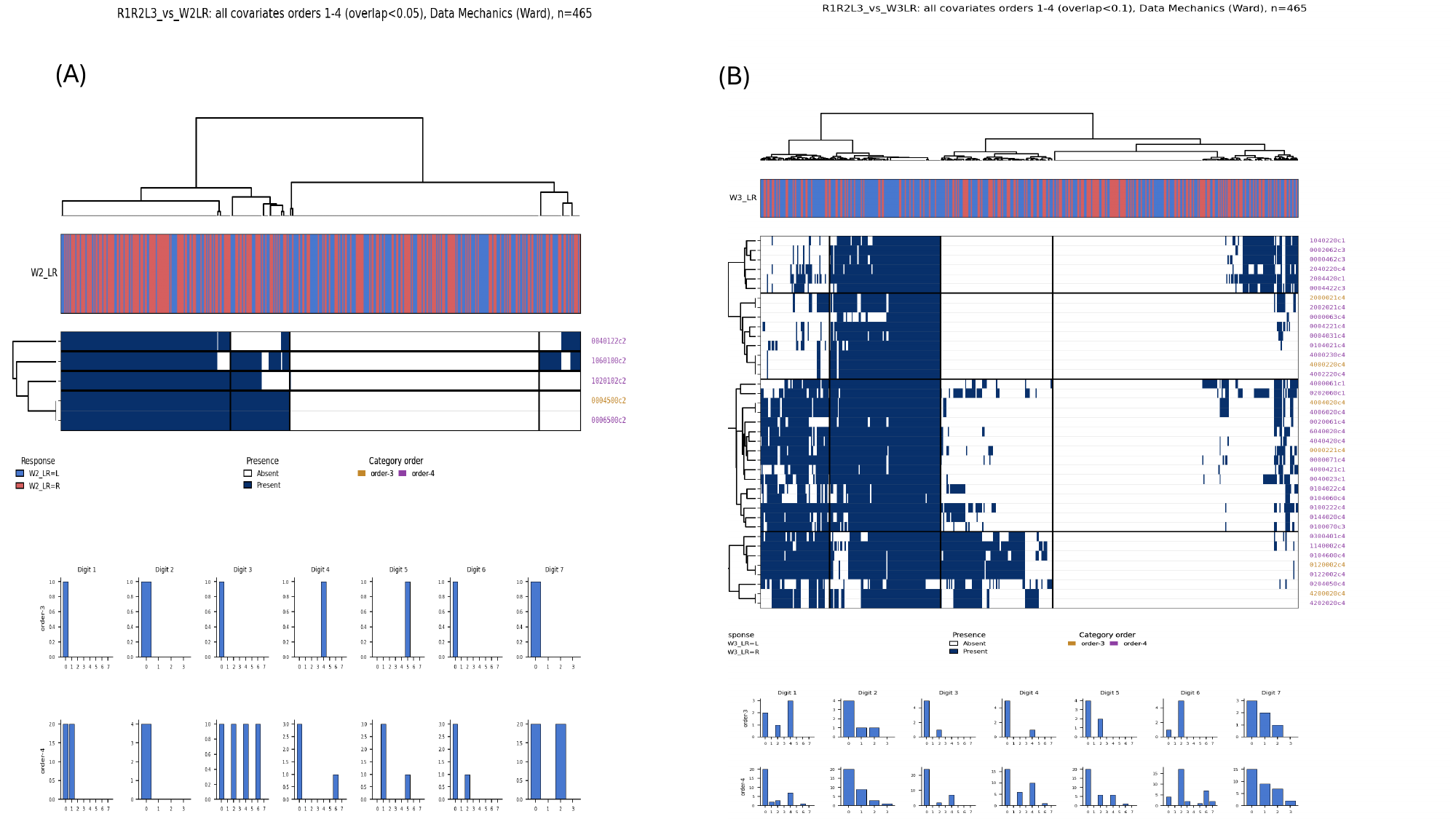}
 \caption{Two heatmaps for splitting R1R2L3 w.r.t (A) W2(3)-vs-W2(4); (B)W3(1\&3)-vs-W3(2\&4)}
 \label{Ttree4thR1R2L3}
 \end{figure}

Thirdly, splitting R1R2L3 with respect to: 1)W2(3)-vs-W2(4); 2) W3(1)-vs-W3(3), are carried out and two corresponding heatmaps are reported in Fig.~\ref{Ttree4thR1R2L3}. Again, patterns of heterogeneity are seen in complex form due to large numbers of soybeans being involved here. For the same reason stated above, we select the splitting by W2(3)-vs-W2(4) to result into sub-branches:R1R2L3L4 and R1R2L3R4. They will be further split in the 5th layer.

\begin{figure}[ht!]
 \centering
\includegraphics[width=1.0\textwidth]{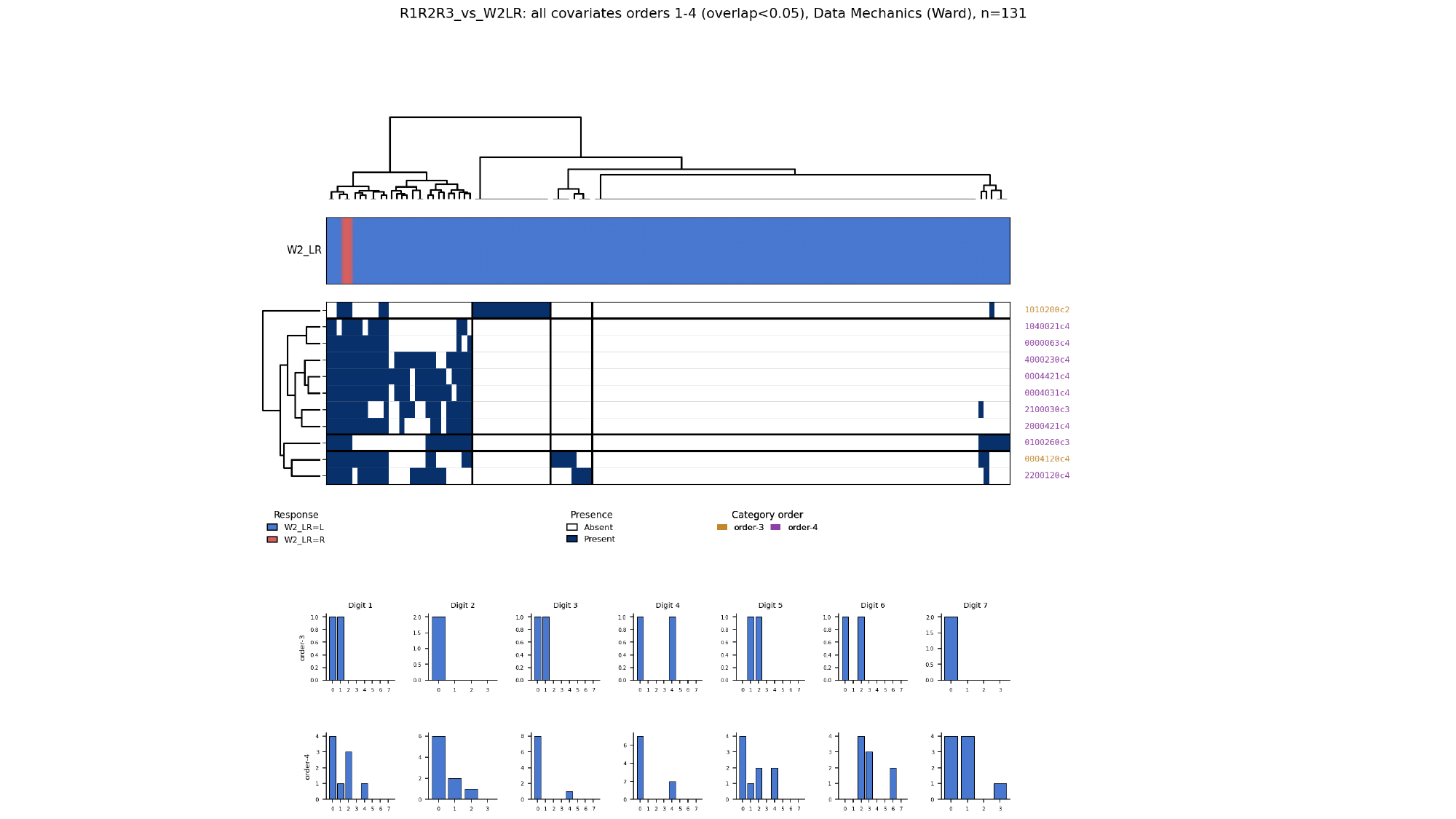}
 \caption{Two heatmaps for splitting R1R2R3 w.r.t W2(3)-vs-W2(4)}
 \label{Ttree4thR1R2R3}
 \end{figure}

Fouthly, the splitting of R1R2R3 with respect to: 1)W2(3)-vs-W2(4); 2) W3(2)-vs-W3(4), are carried out. Only one heatmap is constructed from the corresponding SDA computations because there are no reliable associative information pieces selected for splitting R1R2R3 with respect to W3(2)-vs-W3(4). The heatmap in Fig.~\ref{Ttree4thR1R2R3} indeed indicate no need for splitting with respect to W2(3)-vs-W2(4). That is, R1R2R3 should be kept as an ending node.

\subsection{5th layer}
\begin{figure}[ht!]
 \centering
\includegraphics[width=1.0\textwidth]{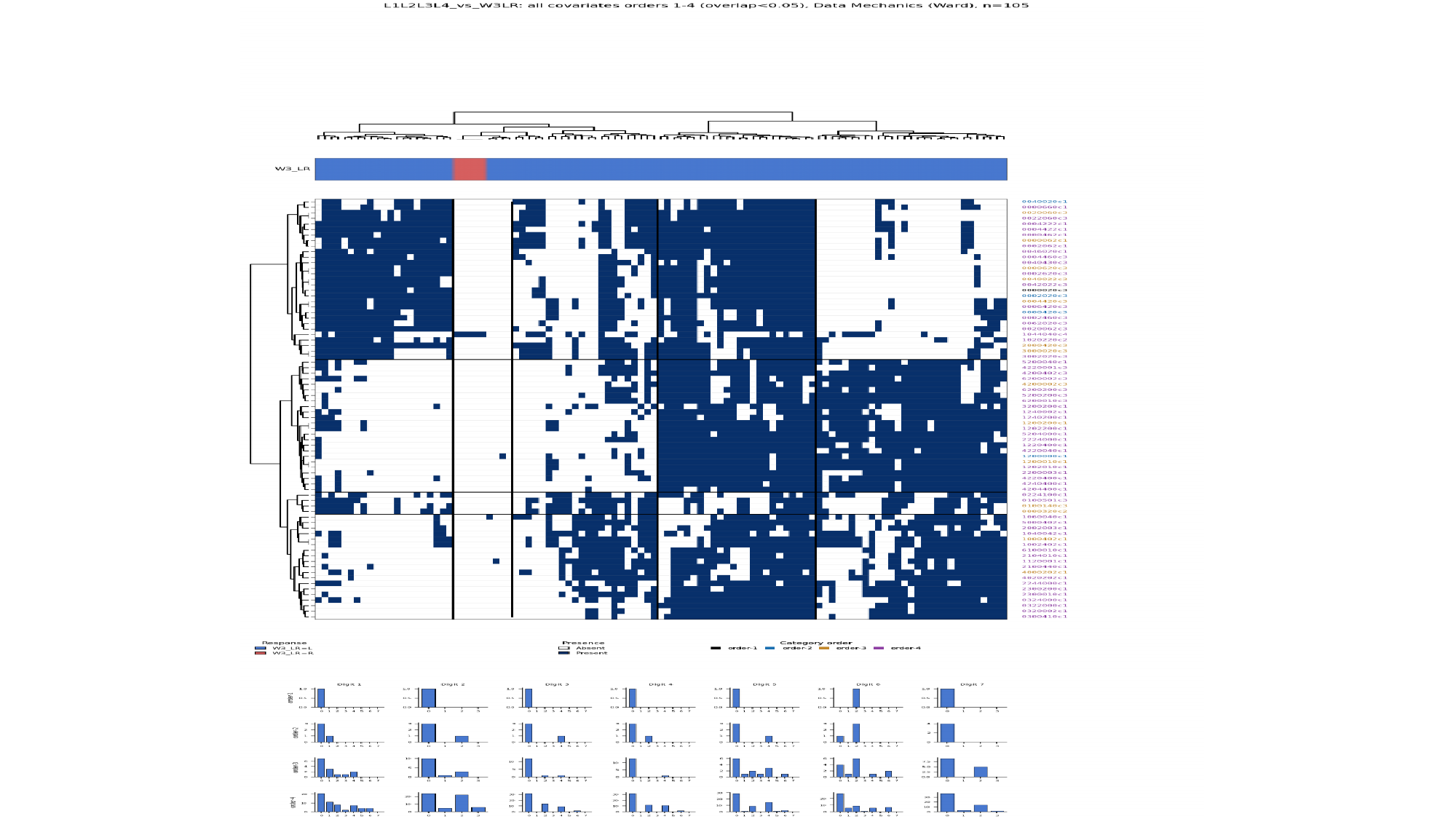}
 \caption{Two heatmaps for splitting L1l2L3L4 w.r.t W3(1)-vs-W3(3)}
 \label{Ttree5thL1L2L3L4}
 \end{figure}

Upon this deepest layer of Taxonomic Hierarchy, the SDA computations on splitting branches: L1L2L3L4 and L1L2L3R4 with respect to W3(1)-vs-W3(3) are carried out to confirm the aforementioned speculation. The computations show no reliable associative information pieces can be collected on the attempt of splitting L1L2L3R4 with respect to W3(1)-vs-W3(3). Though some major feature categories are found on splitting L1L2L3L4 with respect to W3(1)-vs-W3(3), the resultant heatmap, as shown in Fig.~\ref{Ttree5thL1L2L3L4}, indicates the red-vs-blue ratio being too imbalanced. Hence, no splitting on L1L2L3L4 is also determined.

\begin{figure}[ht!]
 \centering
\includegraphics[width=1.0\textwidth]{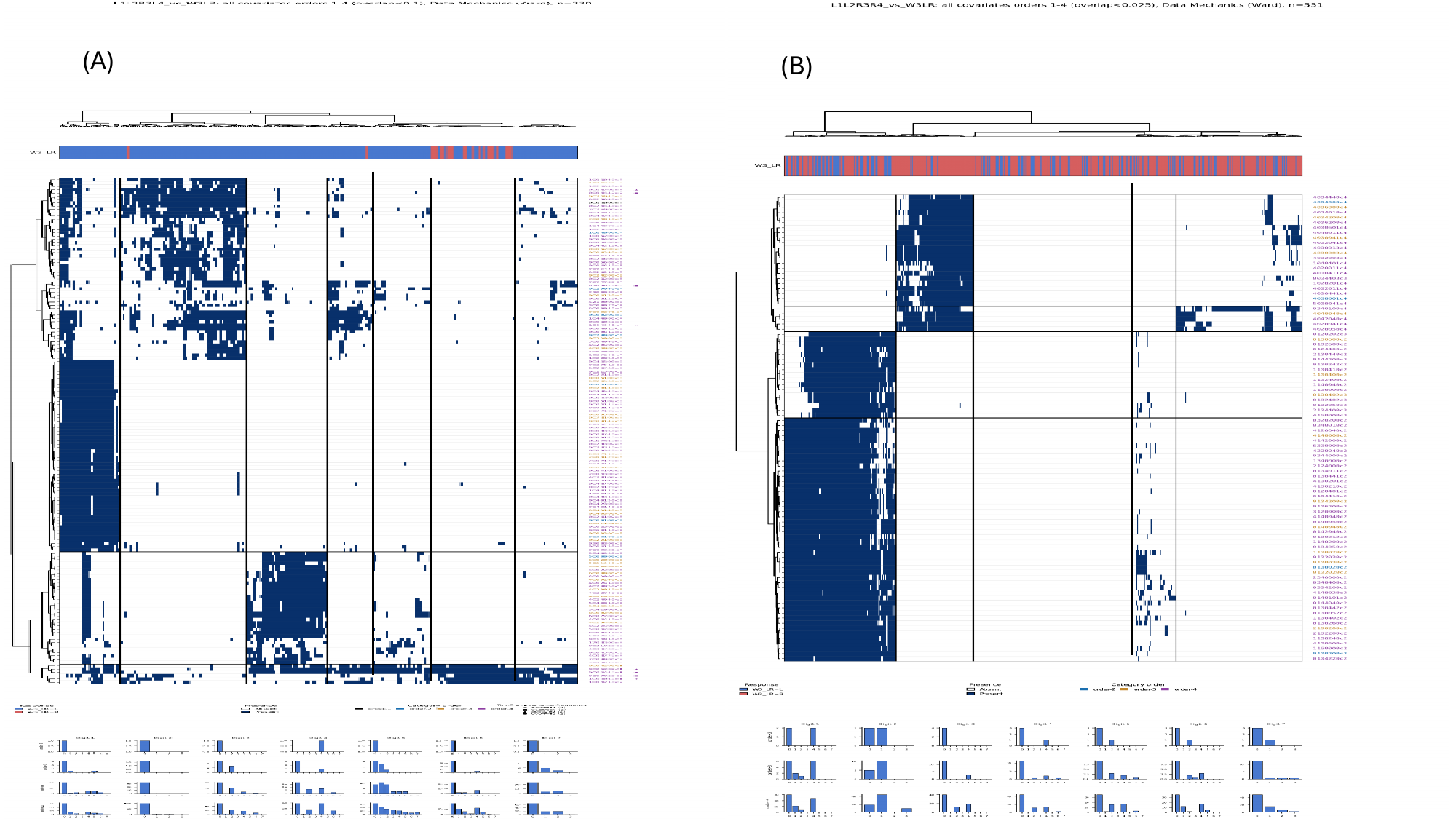}
 \caption{Two heatmaps for splitting L1l2R3L4 (A) and L1l2R3R4 (B) w.r.t W3(1)-vs-W3(3)}
 \label{Ttree5thL1L2R3L4}
 \end{figure}

Further, SDA computations for splitting L1L2R3L4 and L1L2R3R4 with respect to W3(2)-vs-W3(4) are reported and confirmed via the two heatmaps shown in the two panels of Fig.~\ref{Ttree5thL1L2R3L4}. In panel (A), the patterns of heterogeneity support the splitting of L1L2R3L4 into L1L2R3L4L5 and L1L2R3L4R5, while panel (B), the patterns of heterogeneity support the splitting of L1L2R3R4 into L1L2R3R4L5 and L1L2R3R4R5.

\begin{figure}[ht!]
 \centering
\includegraphics[width=1.0\textwidth]{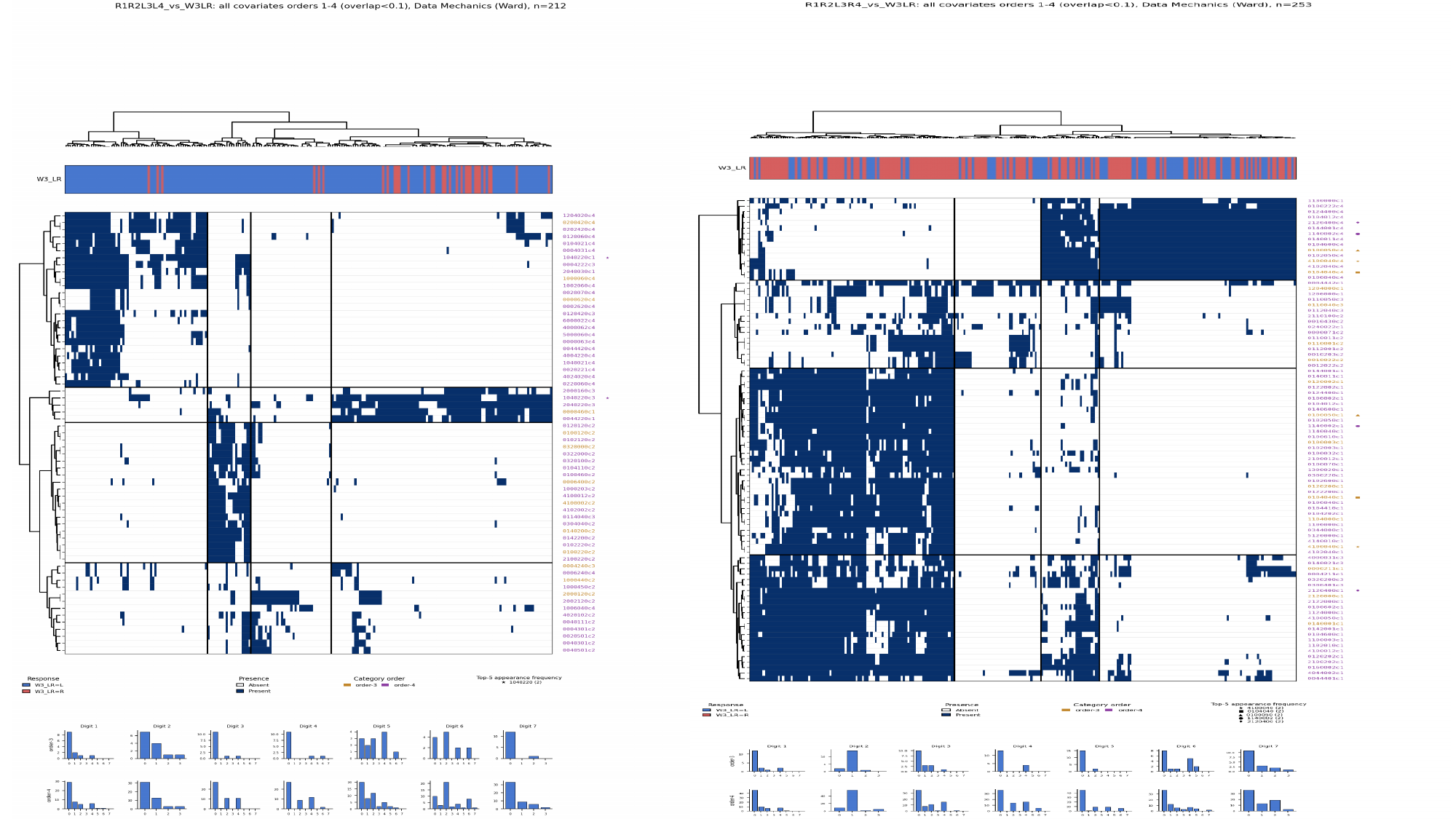}
 \caption{Two heatmaps for splitting R1R2L3L4 (A) and R1R2L3R4 (B) w.r.t W3(1)-vs-W3(3)}
 \label{Ttree5thR1R2L3L4}
 \end{figure}

Furthermore, SDA computations for splitting R1R2L3L4 and R1R2L3R4 with respect to W3(1)-vs-W3(3) are also reported and confirmed via the two heatmaps shown in the two panels of Fig.~\ref{Ttree5thR1R2L3L4}. In panel (A), the patterns of heterogeneity support the splitting of R1R2L3L4 into R1R2L3L4L5 and R1R2L3L4R5, while panel (B), the patterns of heterogeneity also support the splitting of R1R2L3R4 into R1R2L3R4L5 and R1R2L3R4R5.

\subsection{Taxonomic Hierarchy}
\begin{figure}[ht!]
 \centering
\includegraphics[width=1.0\textwidth]{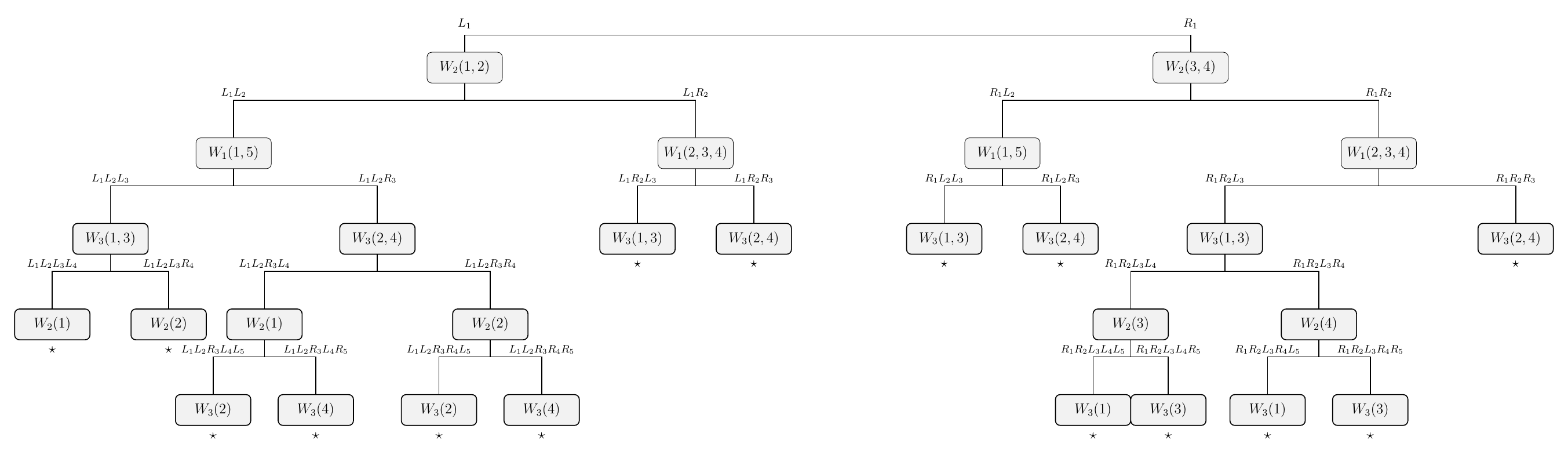}
 \caption{Taxonomic Hierarchy with ending node marked by $\star$.}
 \label{Taxonomictree}
 \end{figure}

\begin{figure}[ht!]
 \centering
\includegraphics[width=1.0\textwidth]{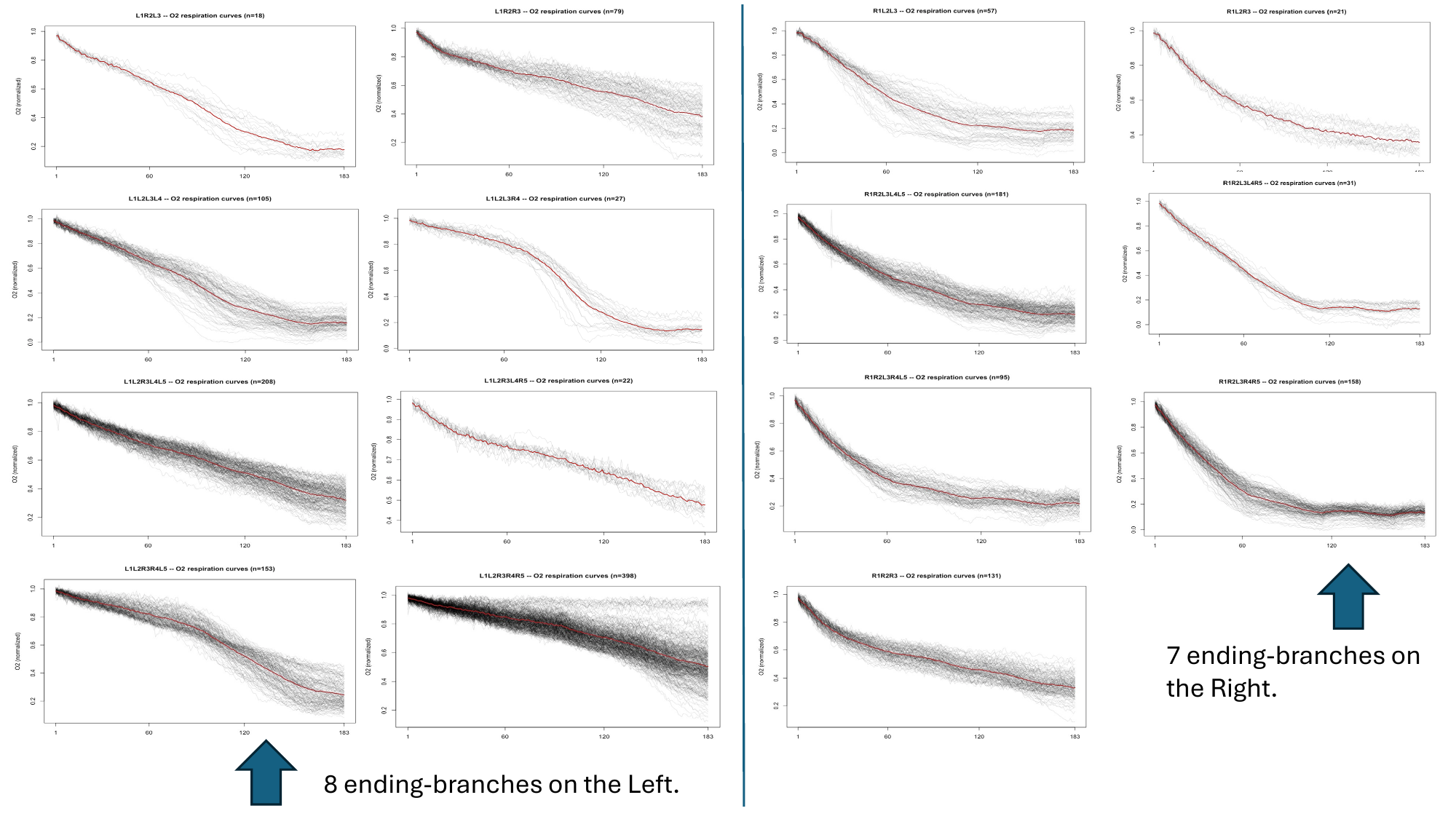}
 \caption{12 typical shapes of O2 respiration curves based on the Taxonomic Hierarchy.}
 \label{12shapes}
 \end{figure}

We finally arrive at a hierarchical binary tree, as shown in Fig.~\ref{Taxonomictree}, with 15 marked ``stars'' as ending nodes. Among these 15 ending nodes, except the one: L1L2R3R4R5, the rest of 14 are reasonable to be typical shapes, as shown in Fig.~\ref{12shapes}. The ending node: L1L2R3R4R5, is seen still embracing shape-heterogeneity. Nonetheless, we stop slitting it because it is an ending node that contains all bad quality soybeans.

Upon Fig.~\ref{12shapes}, it is evident to note that the 8 typical-shapes on the left branch: L1, are closer to the two typical shapes shown in shown in Fig.~\ref{O2Kent} as reported in Fig.1 of \cite{bradford}, while the 7 typical-shapes of exponential decreasing of varying degrees on the right branch: R1, are fundamentally distinct.

Next, we recover all ending nodes' triplet cluster-IDs of $(W1, W2, W3)$, as shown in Table~\ref{tab:W123Cno}. Such triplet cluster-IDs will reveal their triplet of motif-based shapes for better and informative comparisons, which are carried out in the next section.

\begin{table}[h] 
\centering       
\caption{Ending nodes' tree-coordinates and triplet of cluster-\# on Left and Right branches.}
\label{tab:W123Cno}

\begin{tabular}{|l|ccc|ccc|r|} 
\hline
\textbf{Name on Left} & \textbf{W1} & \textbf{W2}& \textbf{W3}& \textbf{W1}& \textbf{W2}& \textbf{W3}& \textbf{Name on Right} \\ \hline
L1L2R3L4L5& 1\&5 & 1& 3& 2\&3\&4& 3& 1& R1R2L3L4L5 \\ \hline
L1L2R3L4R5& 1\&5 & 1& 4& 2\&3\&4& 3& 3& R1R2L3L4R5 \\ \hline
L1L2R3R4L5& 1\&5 & 2& 2& 2\&3\&4& 4& 1& R1R2L3R4L5 \\ \hline
L1L2R3R4R5& 1\&5 & 2& 4& 2\&3\&4& 4& 3& R1R2L3R4R5 \\ \hline
L1L2L3L4& 1\&5 & 1& 1\&3& 2\&3\&4& 3\&4& 2\&4& R1R2R3 \\ \hline
L1L2L3R4& 1\&5 & 2& 1\&3& 1\&5& 3\&4& 1\&3& R1R2R3 \\ \hline
L1R2L3& 2\&3\&4 & 1\&2& 1\&3& 1\&5& 3\&4& 1\&3& R1R2R3 \\ \hline
L1R2R3& 2\&3\&4 & 1\&2& 2\&4& & & & \\ \hline
\end{tabular}

\end{table}

\section{Explanatory exploration}
The Taxonomic Hierarchy shown in Fig.~\ref{Taxonomictree} reveals 15 key shapes of O2-respiration curves as shown in Fig.~\ref{12shapes}. Among these 15 key shapes, there will be $\binom{15}{2} (=105)$ possible pairwise comparisons. It is intuitive that each of these 105 pairwise comparisons will shed lights on one specific aspect of dynamics of Soybean germination. It is worth reiterated that such a comparison is performed computationally based on a corresponding Re-Co dynamics. This Re-Co dynamics is equipped with a binary response variable and all covariate features. By applying the SDA computing, such a Re-Co-dynamics is explored and displayed through a heatmap having a collection of selected major feature-categories of various orders, which are  associative information pieces in the forms of interacting effects of various orders, being arranged along the row-axis and all involving soybean-IDs being arranged along the column-axis. As described through Data Mechanics, both axes of heatmap would undergo iterative permutations via hierarchical clustering algorithm with iteratively defined Euclidean distances to reveal its block-structure.

In this section, we select and report only four pairwise comparisons: one class in L1 branch and another class from R1 branch. These four pairs are selected primarily due to their potentials of biological implications. Though this paper is primarily concerned about the useful exploratory computations for biologists, we still feel obligated to compare a couple of typical O2-respiration shapes, as mentioned and shown in Fig.~\ref{O2Kent} taken from Fig.1 of \cite{bradford}, with a couple of drastically distinct shapes computed in this paper.

Therefore, our primary goal in this section is to first explicitly demonstrate through the heatmap platform: Which covariate feature-categories can precisely provide associative information to efficiently differentiate between identified typical shapes? Secondly, upon each block of 1's within the block-structure sustaining each heatmap, a block-specific network is built to show members of the involving cluster of major feature-categories of various orders and members of the cluster of soybean-IDs. Since such a block framed by these two clusters jointly is indeed a focal point of attention for dynamics underlying the comparison.

\begin{figure}[ht!]
 \centering
\includegraphics[width=1.0\textwidth]{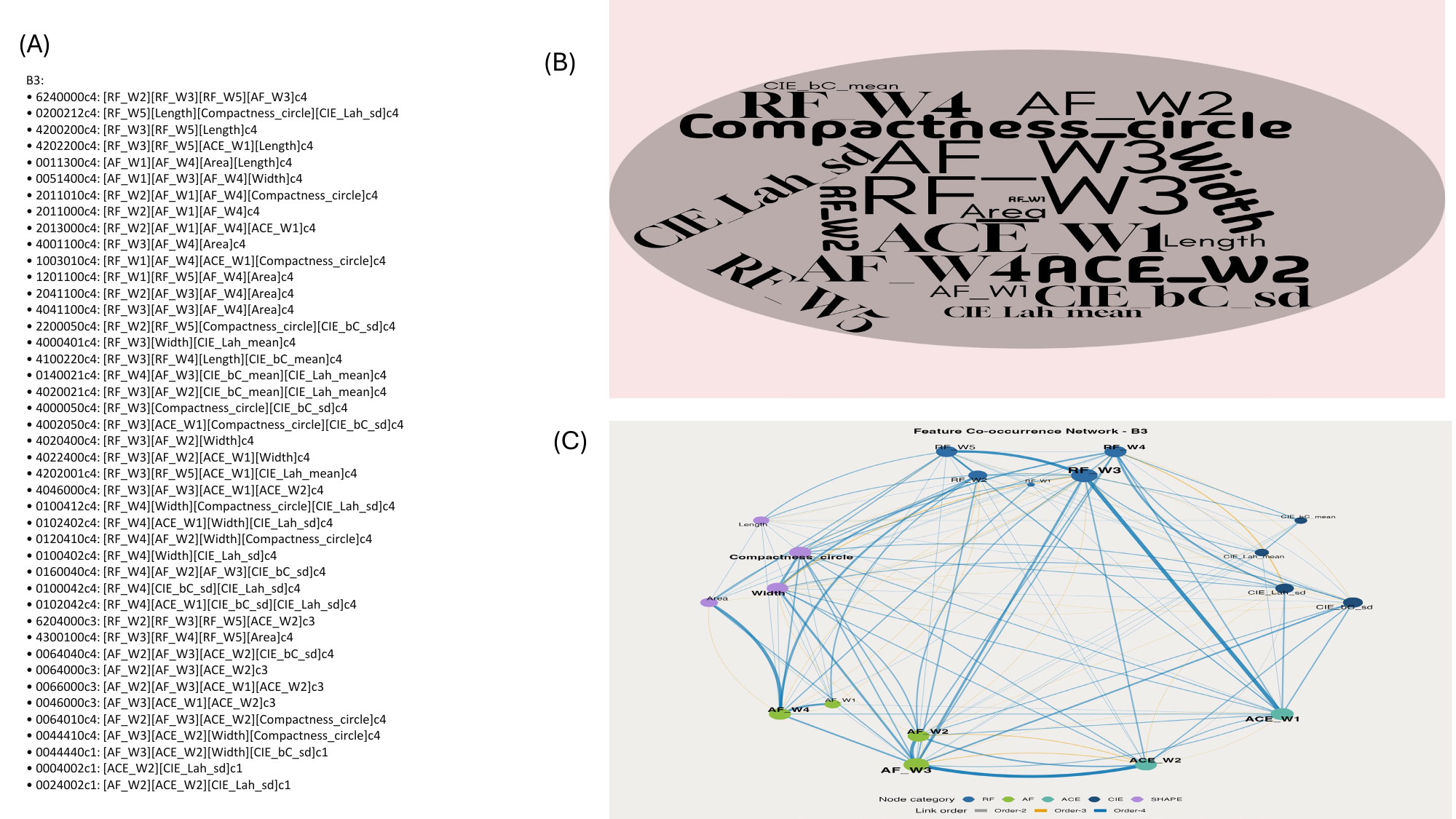}
 \caption{Illustration of a block-specific collection of covariate feature-categories with 7 digit-codes and names (A) and corresponding Word-cloud (B)and network (C).}
 \label{Wordcloud}
 \end{figure}

We recalled here that the 7-digit encoding scheme is carried out with respect to the grouping the 19 covariate features mentioned in the section of Materials and Methods. Each digit is a binary expansion with respect to the presence and absence of covariate features within a group. For instance, a group of covariate features: $(a, b, c)$, with a presence-absence triplet $(1_{a}, 1_{b}, 1_{c})$ is given a digit code: $1_{a}+ 1_{b}\times 2^{1}+ 1_{c}\times 2^2$.
\begin{itemize}
\item[Group 1:] $(RF_{W1}, RF_{W2}, RF_{W3})$;
\item[Group 2:] $(RF_{W4}, RF_{W5})$;
\item[Group 3:] $(AF_{W1}, AF_{W2}, AF_{W3})$;
\item[Group 4:] $(AF_{W4}, ACE_{W1}, ACE_{W2})$;
\item[Group 5:] $(Area, Length, Width)$;
\item[Group 6:] $(Compactness_{Circle}, CIE_{bc-mean}, CIE_{bc-sd})$;
\item[Group 7:] $(CIE_{Lah-mean}, CIE_{Lah-sd})$.
\end{itemize}
See panel (A) of Fig~\ref{Wordcloud} for illustration of a block-specific collection of covariate feature-categories with 7 digit-codes and full names of selected major feature-categories. This list of selected major feature-categories are mechanistically dependent by sharing the same cluster of soybean-IDs. They jointly manifest a component dynamics of the one represented by the whole heatmap. How to read, see and explain this dynamics?

The 7-digit-based histograms, presented at the bottom of heatmap with respect to different orders, are not easy to read, not mentioning to interpret the dynamics content. An improved summarizing approach is ``Word-cloud'', as shown in the panel (B). It reveals which covariate features have high frequencies of appearances. For instance, $RF_{W4}$, $ACE_{W1}$, $AF_{W3}$ appear quite often. In contrast, $RF_{W1}$ is really rare. These summaries are ``marginal order-1 appearances''. The list of selected covariate feature-categories in panel (A) are primary order-4 interacting effects. How to explicitly show such high order interacting relations?

We adopt a special network construction to almost address the above issue fully. A network of 19 nodes is built according to the following protocol.
\begin{itemize}
\item[N1:] A node's disk-size reflects its marginal appearances. A node with order-1 effect has a star-shape, not disk.
\item[N2:] An edge linking between two nodes is colored with respect to a major feature-category's order. Gray for order-2, orange for order-3 and blue for order-4.
\item[N3:] All 19 nodes are fixed on 2D network geometric locations. Members of each group are close to each others.
\end{itemize}
The N3 is required for effective comparisons between and among different networks. According the above network construction protocol, a colored network is illustrated and presented in panel (C) of Fig~\ref{Wordcloud}. Though spelling out details regarding the dynamics embraced by the network in panel (C) is rather a challenging task. Nonetheless, differences between two networks are visible and explainable. This network comparing task is clearly posed in the next four figures below when comparing four pairs of typical shapes of O2-respiration curves.

\begin{figure}[ht!]
 \centering
\includegraphics[width=1.0\textwidth]{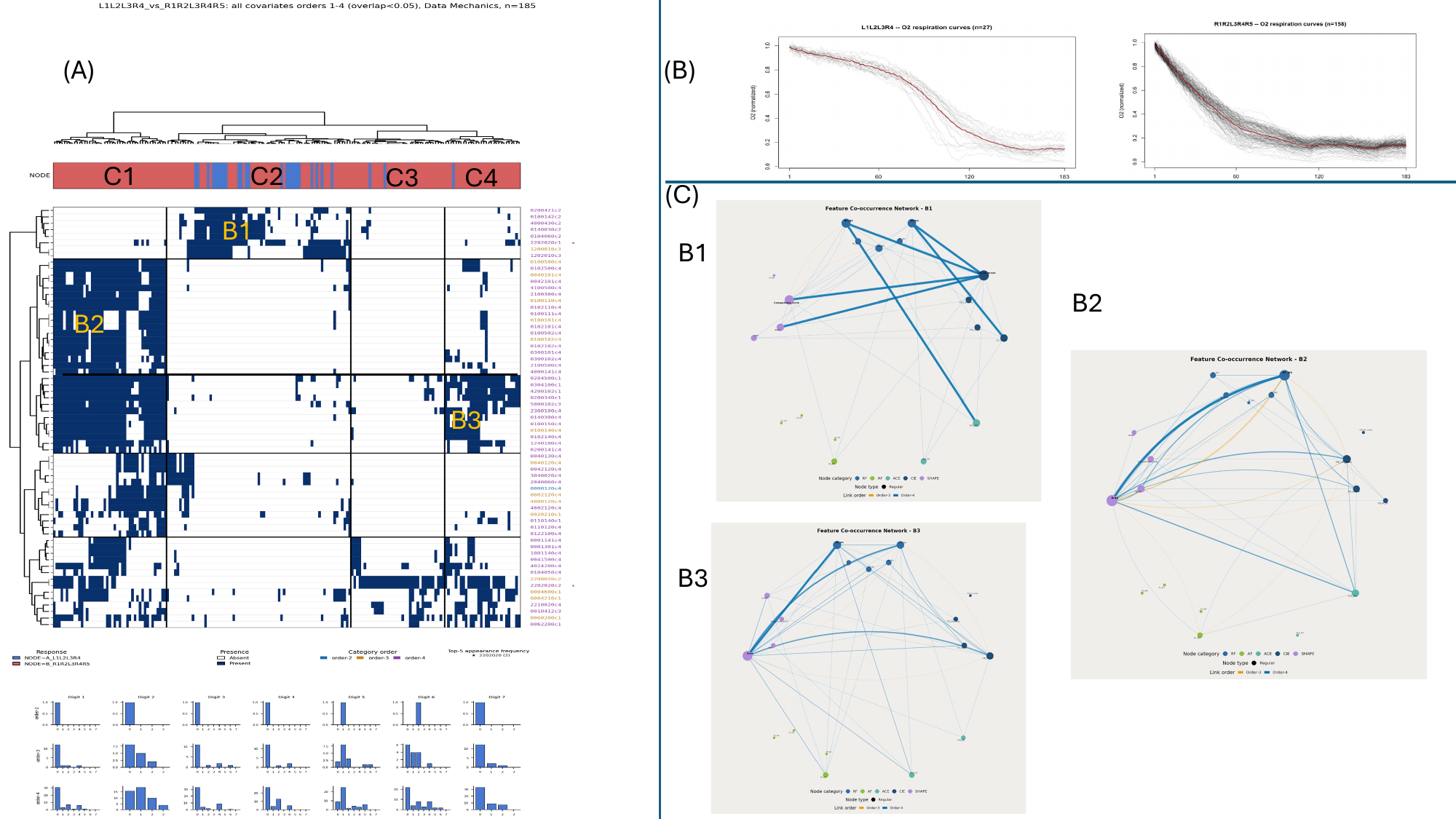}
 \caption{Heatmap for key comparison of two classes:L1L2L3R4-vs-R1R2L3R4R5.}
 \label{comparno1}
 \end{figure}

The first comparison is regarding the two topical shapes: L1L2L3R4-vs-R1R2L3R4R5, as shown in the panel (B) of Fig.~\ref{comparno1}. The shape of O2-respiration of L1L2L3R4 is very close to the faster declining one shown in Fig.~\ref{O2Kent}. It is noted that seeds discussed in \cite{bradford} are much small then soybean. In contrast R1R2L3R4R5 represents the steepest exponential decreasing curve pattern among the 7 typical-shapes on the right branch: R1. This comparison is meant to show the dynamics that sustaining the differences between two ``extreme'' cases.

The resultant heatmap, as shown in panel (A), has an evident block structure. On its column-axis, 4 clusters of soybean-IDs are marked and encoded: C1 to C4, under the HC-tree, while five clusters of major feature-categories partition the row-axis of 7-digit-code-names of selected feature-categories. Within the $5\times 4$ block-structure, three blocks of primary 1's are marked: B1, B2 and B3. The block-B1 is framed by soybean-ID cluster C2, which contains majority of members of L1L2L3R4 (in blue), while block-B2 and block-B3 are purely members of R1R2L3R4L5 (in red). As such, cluster-C2 is characterized by soybean-IDs which are mostly participating major feature-categories belonging to 1st row-cluster framing the block-B1, and simultaneously being absent from all major feature-categories belonging to  2nd and 3rd row-clusters framing block-B2 and block-B3, respectively.

It is evident to see the differences and similarity among the three networks built according to 1st, 2nd and 3rd row-clusters of major feature-categories. Particularly in panel (C), we can evidently see that the network-B1 is drastically distinct from network-B2 and network-B3. Network-B1 has 6 ``thick-edges'', which likely play the role of key factors of order-2 to order-4, that are missing completely missing in the network-B2 and network-B3. These 6 thick edges are linking among 7 covariate-features:
\[
\{RF_{W4}, RF_{W4}, ACE_{W1}, Compactness_{Circle}, Width, CIE_{Lah-mean}, CIE_{Lah-sd}\}
\]
These key factors are supposed to characterize cluster-C2, so to characterize the typical shape of L1L2L3R4 when comparing with R1R2L3R4R5. In contrast, the main and secondary linkages of network-B3 and network-B2, like the main thick-edge between $RF_{W4}$ and $Area$, seem to overlap with to a great extent.

\begin{figure}[ht!]
 \centering
\includegraphics[width=1.0\textwidth]{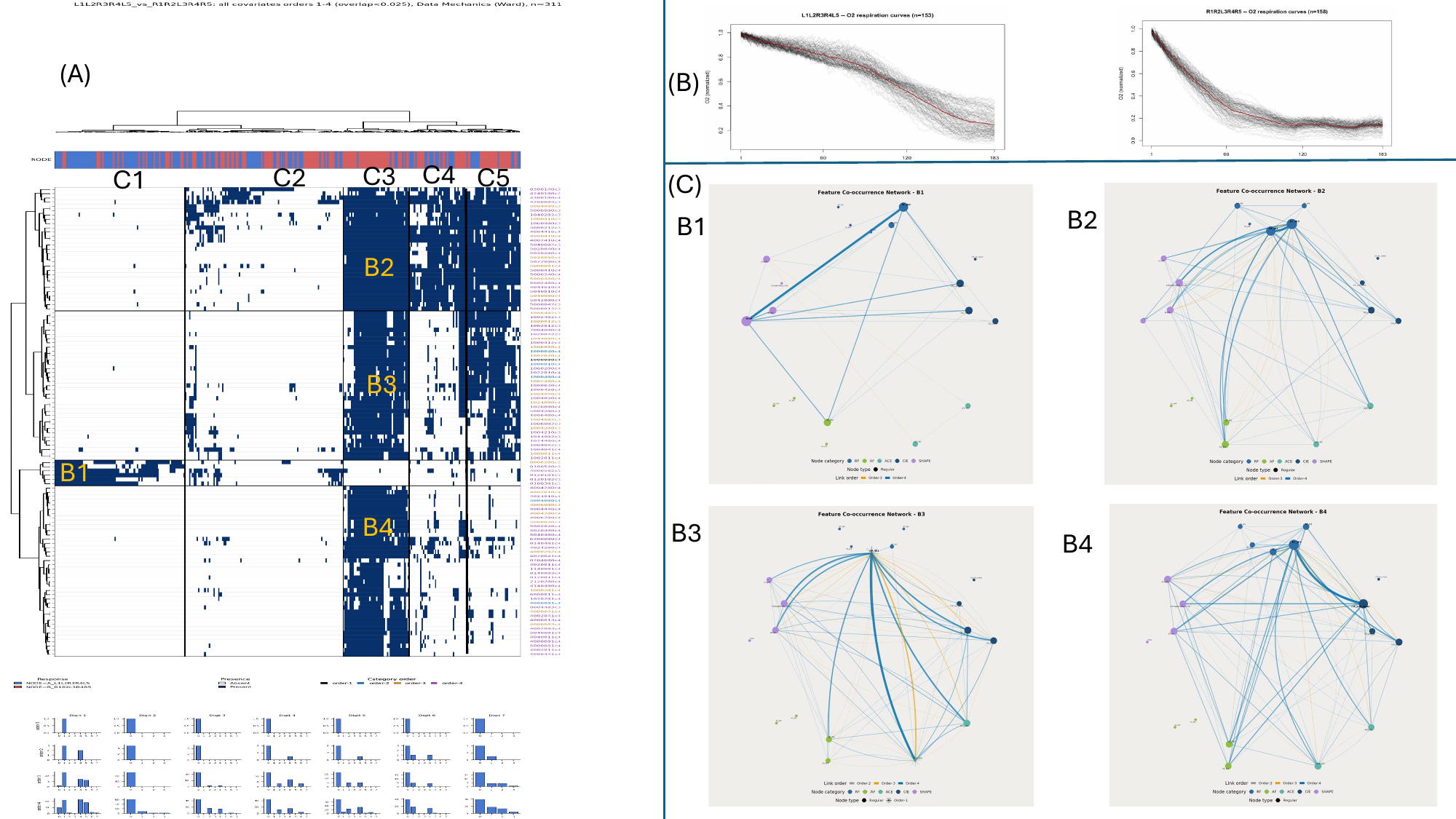}
 \caption{Heatmap for key comparison of two classes:L1L2R3R4L5-vs-R1R2L3R4R5.}
 \label{comparno2}
 \end{figure}

The second comparison is regarding the two topical shapes: L1L2R3R4L5-vs-R1R2L3R4R5, as shown in the panel (B) of Fig.~\ref{comparno2}. The shape of O2-respiration of L1L2R3R4L5 indicates a quality being slightly worse than that of L1L2L3R4, which are very close to the slower declining one in Fig.~\ref{O2Kent}. The heatmap in panel (A) is sustained by a $4\times 5$ block-structure, which are framed by 5 column-clusters of soybean-IDs and 4 row-clusters of major feature-categories. Among these 20 block, four blocks of almost 1's are marked: B1 to B4.

Upon the block-B1, its cluster-C1 consists primarily members of L1L2R3R4L5, and its 1st row-cluster is represented by the B1-network in panel(C). It is interesting to note that network-B1 has only one thick-edge between $RF_{W4}$ and $Area$. From this thick-edge perspective, this B1-network is structurally distinct from that of network-B2, network-B3 and network-B4, which are relatively similar. This similarity is clearly recognized via thick-edges linking $RF_{W1}$ and $RF_{W3}$ to many other covariate features. Both $RF_{W1}$ and $RF_{W3}$ are relatively small nodes in network-B1. It is specifically rare that $RF_{W1}$ is indeed an order-1 factor in network-B3, so it is marked with a ``star'', instead of disk. Finally, we speculate that the similarity of these three network implies three distinct phases of one highly important dynamics.

Finally, we mention one significant distinction of  network-B2, network-B3 and network-B4 in panel (C) of Fig.~\ref{comparno2} from the two: network-B2 and network-B3 in panel (C) of Fig.~\ref{comparno1}, is the heavy involvements of $RF_{W1}$ and $RF_{W3}$, rather than $RF_{W4}$. The biological implications from multi-spectral imaging perspective are likely important.

\begin{figure}[ht!]
 \centering
\includegraphics[width=1.0\textwidth]{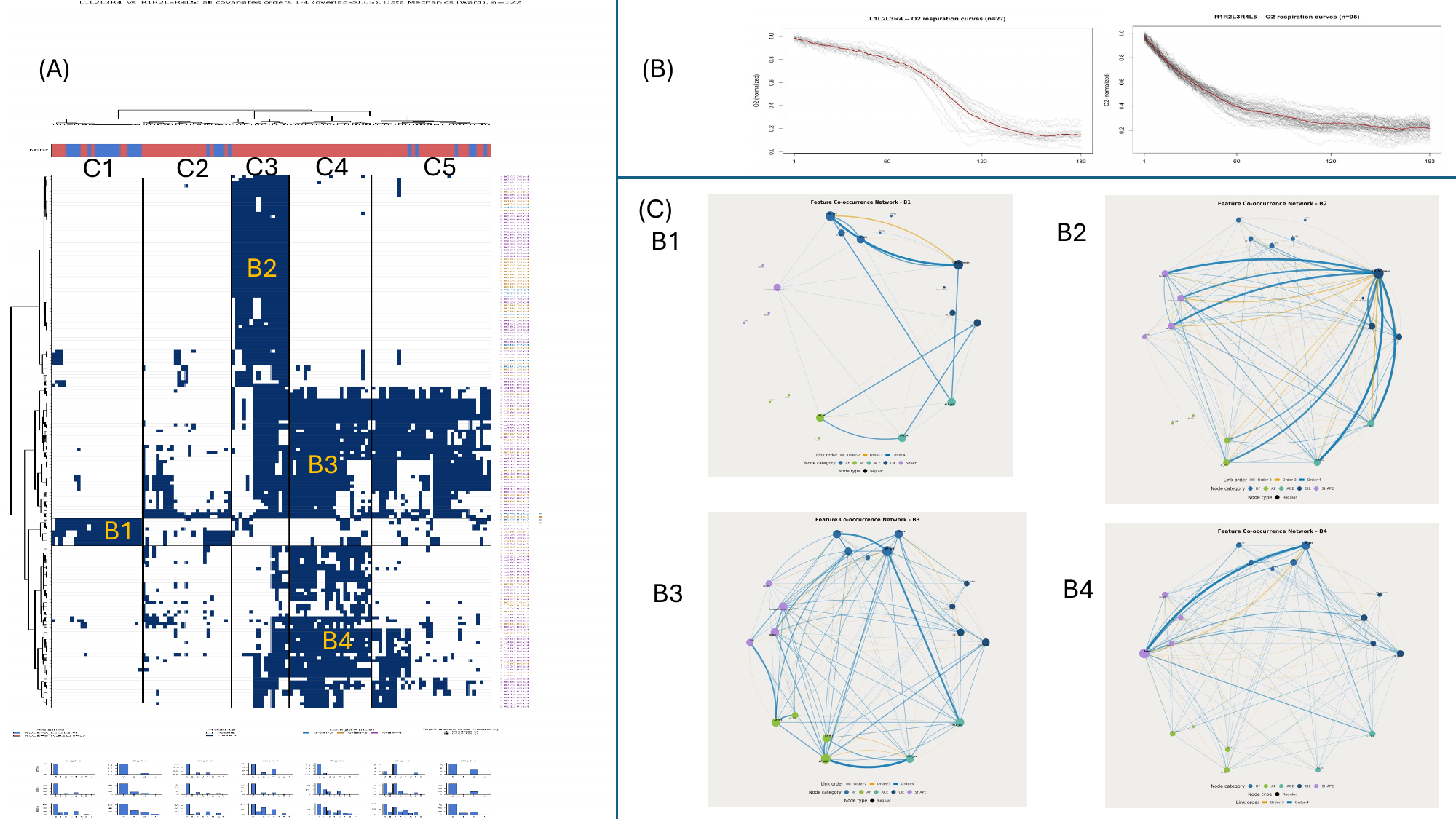}
 \caption{Heatmap for key comparison of two classes:L1L2L3R4-vs-R1R2L3R4L5.}
 \label{comparno3}
 \end{figure}

The third comparison is regarding the two topical shapes: L1L2L3R4-vs-R1R2L3R4L5, as shown in the panel (B) of Fig.~\ref{comparno3}. This comparison is meant to contrast with the 1st comparison. Again L1L2L3R4 is a typical shape for the best quality ones, while R1R2L3R4L5 also represents an exponential decreasing shape, but its decreasing speed is not as extreme as that of R1R2L3R4R5.

The heatmap computed for such a comparison, as shown in the panel (A) of Fig.~\ref{comparno3}, is sustained by a $4\times 5$ block-structure. Among the 20 blocks, four blocks: B1 to B4, are marked. It is clear to see that all 5 clusters of soybean-IDs can be characterized by presences or absences across the four blocks' row-clusters of major feature-categories. For instance, cluster-C1, which contains the majority of members of L1L2L3R4, is characterized by a vertical block-chain: the block-B1 of almost 1's and three blocks of 0's corresponding with the rest of row-clusters. Likewise, the rest of 4 clusters of soybean-IDs are characterized.

The four row-clusters with respect to block-B1 through block-B4 are built into four networks, as shown in the panel (C) of Fig.~\ref{comparno3}. The network-B1 is relative simple by having only thick-edge between $RF_{W1}$ and $CIE_{Lah-mean}$. This observed unique thick-edge stands structural distinction from the rest of three networks. These three networks: B2 to B4, are clearly much more complex that networks shown in panel (C) of Fig.~\ref{comparno1} and Fig.~\ref{comparno2}. They have a wider range of hubs and simultaneously have many more thick-edges. Such differences in complexity might bear with biological significance.

\begin{figure}[ht!]
 \centering
\includegraphics[width=1.0\textwidth]{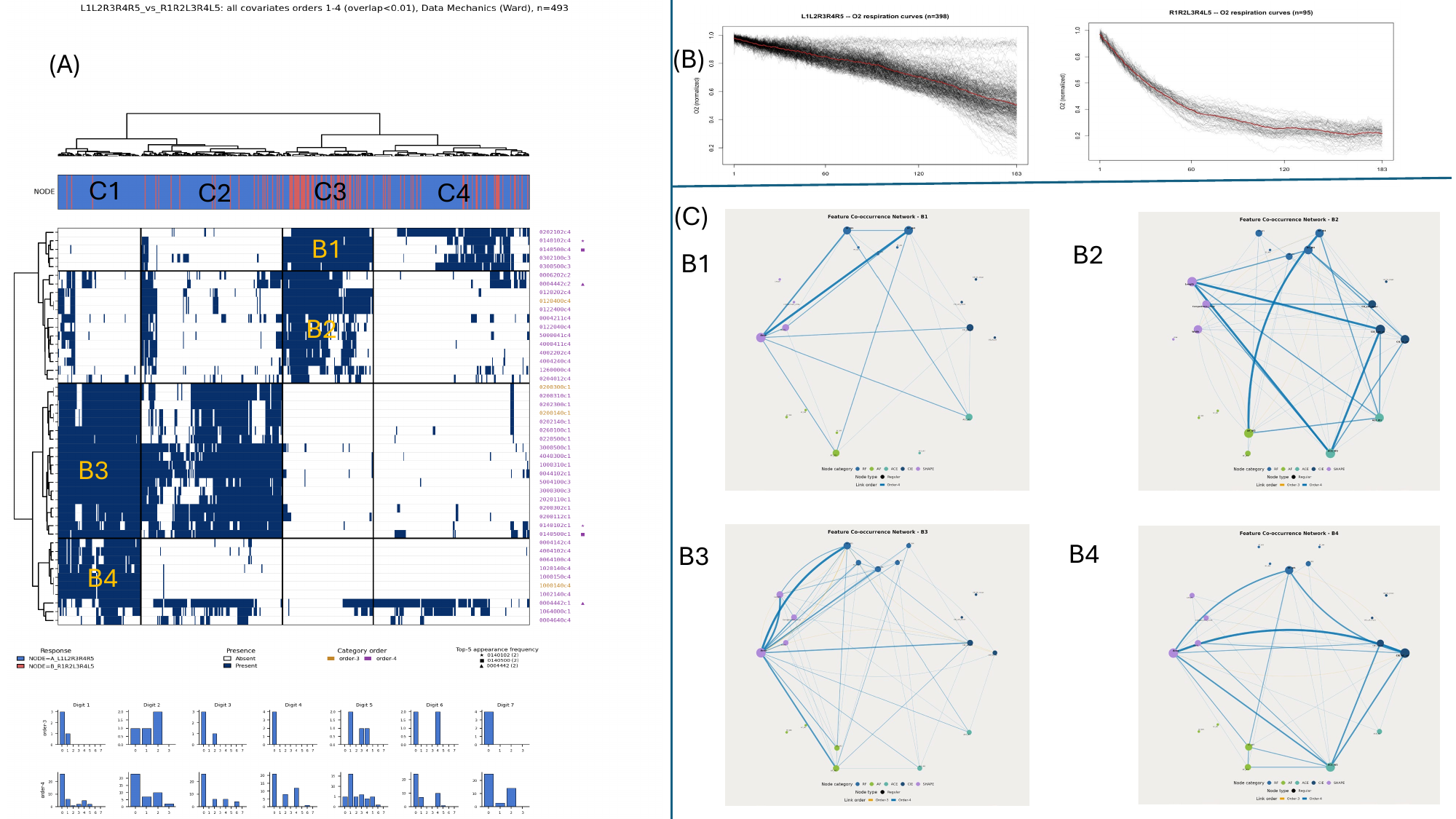}
 \caption{Heatmap for key comparison of two classes:L1L2R3R4R5-vs-R1R2L3R4L5.}
 \label{comparno4}
 \end{figure}

The 4th comparison is regarding the two typical shapes: L1L2R3R4R5-vs-R1R2L3R4L5, as shown in the panel (B) of Fig.~\ref{comparno4}. The L1L2R3R4R5 represents the worse quality of soybean. The heatmap of this comparison, as shown in panel (A), is found being sustained by a $4\times 4$ block-structure. Among all 16 blocks, four blocks are marked. The 4 clusters of soybean-IDs are characterized by their vertical block-chains, respectively. In particular, cluster-C3, which primarily consists of members of R1R2L3R4L5, is characterized by the B1 and B2, which are almost 1's,  and empty-counterparts of B3 and B4.

The four networks show four distinct patterns. However, when pooling characteristics of network-B1 and network-B2 together, we see many more hubs and thick-edges than combination of network-B3 and network-B4. This observational pattern highlights typical shape of R1R2L3R4L5 in comparing with L1L2R3R4R5, within which likely contains many dead soybeans having horizontal O2-respiration curves.

Both network-B3 and network-B4 share a common hub: $Area$. However, the two sets of thick-edges in these two network are linking $Area$ to two rather distinct set of big nodes. Such differences are strikingly pointing to existing heterogeneity with L1L2R3R4R5.

\section{Conclusion}
In this paper, we develop Computational Taxonomy to resolve two central issues in soybean related seed science. First issue is how to extract and categorize typical-shapes of O2-respiration curves? The second issue is how to discover and explain the dynamics underlying any comparison between two typical-shapes? Our data-driven computational developments for resolving these two issues are to primarily help biologists in this soybean related research field. Nonetheless, this methodology is indeed essential to sciences beyond seed research.

In dealing with functional curves embracing with categorical typical-shape coupled with varying variations along the temporal axis, the development of discovering motifs turns out to be critical from scientific and practical perspectives. The three fused variables based on three resultant motifs: (W1, W2, W3), provide critical insights into a wide spectrum of typical-shapes among the 1684 O2-respiration curves.

Then, the Taxonomic Hierarchy computations dynamically group separated, but similar typical-shapes together to significantly reduce the complexity of the combinatorial spectrum of typical-shapes. This reduction is based on SDA computed associative information pieces and their collective display via heatmap. Every heatmap is shown being sustained by a block-structure, which offers multiple horizontal block-chains to characterize multiple component-specific dynamics via network, and simultaneously offers multiple vertical block-chains to characterize heterogeneity among involved soybean-IDs.

Very importantly, the Taxonomic Hierarchy's ending nodes are the finally resultant typical-shapes. This set of typical-shapes collectively shown the heterogeneity-vs-homogeneity map. That is, such a computed typical shape is a homogeneity-locality, which is explicitly expressing what kind of randomness is embraced at this locality.  With respect to this map, any comparison of typical-shapes becomes finite-sample meaningful.

In our future research, we plan to address the scientific issue: How to compose all comparison-based dynamics into one whole dynamics underlying soybean's O2 respiration curves contained in this data set? We believe that soybean related science has a solid foundation based on Computational Taxonomy presented in this paper. Also, we hope to expand and to see its concept and computing being shown applicable in a wide spectrum of scientific fields in the near future.

\bibliographystyle{unsrt}

\end{document}